   \documentclass{aa}
   \usepackage{graphicx}
   
\begin{document}
   \title{The structure of Planetary Nebulae: theory vs. practice. 
 \thanks{Based on observations made with ESO Telescopes at the La Silla 
Observatories (program ID 65.I-0524), 
and TNG (Telescopio Nazionale Galileo) at La Palma, Canary Islands (program AOT10-10). 
%and the NASA/ESA Hubble Space Telescope, 
%obtained from the data archive at the Space Telescope Institute. 
%Observing programs: GO 7501 and GO 8773 (P.I. Arsen Hajian). 
%STScI is operated by the association of Universities for Research in   
%&Astronomy, Inc. under the NASA contract  NAS 5-26555.  
We extensively 
apply the photo--ionization code CLOUDY, developed at the Institute of
Astronomy of the Cambridge University (Ferland et al. 1998).}
 \author{ F. Sabbadin  \inst{1} \and M. Turatto \inst{1} \and R. Ragazzoni \inst{2} \and E. Cappellaro \inst{3} \and S. Benetti \inst{1} }
   \offprints{F. Sabbadin, sabbadin@pd.astro.it}
   \institute{INAF - Astronomical Observatory of Padua, Vicolo dell'Osservatorio 5, I-35122 Padua, Italy \and  
   INAF - Astrophysical Observatory of Arcetri, Largo E. Fermi 5, I-50125, Italy \and 
   INAF - Astronomical Observatory of Capodimonte, Via Moiariello 11, I-80131 Naples, Italy  
  %\and Max-Planck-Institut f\"ur Astronomie, Koenigstuhl 17, D-69117 Heidelberg, Germany
}}
   \date{Received November 21, 2005; accepted January 17, 2010}
   \abstract{This paper - the first of a short series dedicated to the long-standing astronomical problem of de-projecting the bi-dimensional apparent morphology of a 
three-dimensional mass of gas - focuses on the density distribution in real Planetary Nebulae (and all   
types of expanding nebulae). We  
introduce some basic theoretical notions, discuss the observational methodology and develope the accurate procedure for the  determination of 
the matter radial profile within the sharp portion of nebula in the plane of the sky identified by the zero-velocity-pixel-column (zvpc) of high-resolution 
spectral images. 
Moreover, a series of 
evolutive snapshots is presented, combining illustrative examples of model- and true-Planetary Nebulae. Last, the general and specific applications of the method 
(and some caveats) are discussed.
   }
   
   \titlerunning{Planetary Nebulae: theory vs. practice }
   
   \maketitle
%
%________________________________________________________________
\section{Introduction} 

The spatial distribution of gas constitutes the most important observational parameter for Planetary Nebulae (PNe): when combined with gas 
dynamics, it 
supplies fundamental information on the mechanisms and physical processes driving nebular evolution (mass-loss history, wind interaction, ionization, 
magnetic fields, binarity of the central star etc.) and allows us a direct comparison of each real object with theoretical evolutionary models, 
hydro-dynamical simulations and photo-ionization codes.

In spite of this, the whole astronomical literature dedicated to PNe contains only a handful of papers dealing with the true spatial 
structure, which is  
generally obtained by deconvolution of volume emissivity from the apparent image using Abel's integration equation for spherically 
symmetric systems and ``ad-hoc'' algorithms for inclined axi-symmetrical nebulae (Wilson \& Aller 1951, Lucy 1974, Soker et al.
1992, Volk \& Leahy 1993, Bremer 1995). These rare - quite rough and controversial - results  stress  the huge difficulties so far encountered in 
de-projecting the bi-dimensional appearance of a three-dimensional structure (Aller 1994).

Recently, Sabbadin et al. (2005 and references therein) have suggested 
that the dynamical properties of the ionized gas represent the key for overcoming the 
stumbling block of de-projection in PNe and other 
types of expanding nebulae, like Nova and Supernova Remnants, shells around Population I Wolf-Rayet stars, 
nebulae ejected by Symbiotic Stars, bubbles surrounding early spectral-type Main Sequence stars etc..

Historically, the general kinematical rule for the bright main shell of PNe goes back to Wilson (1950), based on Coud\'e spectra - no image-derotator - of 26  targets: 
the high-excitation 
zones expand more slowly than the low-excitation ones, and there is a direct correlation between the expansion velocity and 
the size of the monochromatic image. 

Weedman (1968) secured Coud\'e spectra (+ image-derotator) of a sub-set of Wilson's list (10 PNe), covered along the apparent 
major axis. Assuming ``a priori'' the prolate spheroid hypothesis with a/b=1.5, Weedman derived a typical expansion law of the 
type $V_{\rm exp}$=s(R-R$_0$), where s is the (positive) slope of the expansion velocity gradient and R$_0$ the radius at which $V_{\rm exp}$=0. 
In most cases R$_0\simeq$0 or R$_0$$<<$R$_{\rm neb}$ (i.e. the outflow is nearly ballistic, Hubble-type).

The classical papers by Wilson (1950) and Weedman (1968) belong to the heroic age of photographic plates. Later on, the introduction of linear 2-D 
detectors at high efficiency (CCDs) greatly enhanced flux accuracy in a wide spectral range, and the use of echelle spectrographs allowed the observers to extend the 
analysis to large, faint objects. Even so, a odd dichotomy marks out the whole astronomical literature: high-resolution spectra of PNe are used to obtain 
either the detailed kinematics in a few ions (generally,  [O III], H I and [N II]) or the ``average'' nebular fluxes. 

In the former case 
(i. e. kinematics), long-slit observations are performed through an interference filter to isolate a single order containing one or two 
nebular emissions. 
Valuable examples concern the kinematics of: 15 multiple shell PNe (Guerrero et al. 1998; single position angle (PA), instrumental spectral 
resolution $\Delta$V=6.5 km s$^{-1}$), NGC 2438 (Corradi et al. 2000; single PA, $\Delta$V=4.3 km s$^{-1}$) and MZ 3 (a symbiotic-star nebula), covered at 
various PA by 
Santander-Garcia et al. (2004, $\Delta$V=6.0 km s$^{-1}$) and Guerrero et al. (2004, $\Delta$V=8.0 km s$^{-1}$). We must notice, however, that the standard  
observational 
procedure - strongly reducing the useful spectral range by means of an interference filter - represents an excessive precaution  
in most cases, since 
the PN is an emission line 
object. As shown by Benetti et al. (2003) and Sabbadin et al. (2004), a PN larger (even much larger) than the 
order separation of the echellograms can be covered with a single exposure in the whole spectral range, the only limit 
(variable from one instrument to an other)  being  
the effective superposition of spectral images belonging to different orders  (and not the mere order separation). 

In the latter dichotomous case (average fluxes from high-dispersion spectroscopy), line intensities are integrated over the entire slit (whose length is smaller than 
order separation) and provide ``mean'' 
physical conditions and ionic 
and chemical abundances within the whole slice of nebula covered by the spectrograph, thus losing the detailed (i. e. pixel-to-pixel) information 
on kinematics and flux. Representative examples are given by Hyung \& Aller (1998) and 
Hyung et al. (2001). 

A courageous effort for overcoming the foregoing impasse has been performed by Gesicki \& Zijlstra (2000) and Gesicki et al. (2003 and references therein): 
they secured high-resolution 
spectra of a number of compact PNe (including Galactic Bulge objects and nebulae in the Sagittarius Galaxy and in the Magellanic Clouds) 
and combined integrated emission profiles of forbidden and recombination lines with the photo-ionization code of a spherical 
nebula (Torum model), inferring that acceleration is a quite common property in the PNe innermost layers (i. e. ``U''-shaped expansion profile), 
due to the dynamical contribution by the shocked, hot wind from the central star. 

Unfortunately, a detailed comparative analysis (Sabbadin et al. 2005) 
proves that the ``U''-shaped expansion profile is a spurious, incorrect result caused by a combination of unsuited assumptions, 
spatial resolution and diagnostic choice. 

All this confirms that, due to the nebular complexity and large stratification of the radiation and kinematics, the recovery of the spatio-kinematical 
structure needs observations: 

(a) at adequate ``relative''  spatial (SS) and spectral (RR) resolutions (SS=R/$\Delta$r, R=apparent radius, 
$\Delta$r=seeing+guiding; RR=$V_{\rm exp}$/$\Delta$V, $\Delta$V=instrumental spectral resolution),  

(b) in a wide spectral range 
containing emissions at low, medium and high ionization and 

(c) at several slit positions over the nebula,  

to be combined with a straightforward and versatile method of analysis.

To this end, in 2000 we started a multi-PA spectroscopic survey of bright PNe in both hemispheres, observed  with ESO NTT+EMMI (80 echelle orders 
covering the spectral range
$\lambda\lambda$3900-8000 $\rm\AA\/$ with $\Delta$V=5.0 km s$^{-1}$) and Telescopio Nazionale Galileo (TNG)+SARG (54 orders, spectral range
$\lambda\lambda$4600-8000 $\rm\AA\/$, $\Delta$V=2.63 km s$^{-1}$). The pixel-to-pixel analysis of flux and velocity in a large number of emissions 
gives the bi-dimensional ionic structure (tomography) of each nebular slice intercepted by the spectrograph slit and a 3-D rendering 
program, assembling all tomographic maps, provides the accurate spatial distribution of the kinematics, physical 
conditions ($T_{\rm e}$ and $N_{\rm e}$) and ionic and chemical abundances within the nebula. 

Besides the - in fieri - detailed spatio-kinematical study of individual PNe (Ragazzoni et al. 2001, Turatto et al. 2002, Benetti et al. 2003, 
Sabbadin et al. 2004, 2005), we decided to deepen a general, long-lasting open problem: inferring the radial density profile in expanding nebulae. 

This introductory, didactic paper contains the detailed procedure for the determination of the gas distribution in real PNe and is structured as follows: 
Sect. 2 gives some basic theoretical notions, Sect. 3 describes the practical application to  high-dispersion spectra, Sect. 4 presents a 
series of evolutionary snapshots for model- and true-PNe, Sect. 5 contains the general discussion and Sect. 6 draws the conclusions. 

In a future, applicatory paper we 
will compare the density profiles observed in a representative sample of targets with the expectations coming from 
recent hydro-dynamical 
simulations and theoretical models, in order to disentangle the evolutive phenomenology and physical processes responsible for PNe shape and shaping.
\section{Theory}
Following Aller (1984), Pottasch (1984) and Osterbrock (1989), the absolute flux emitted in the line $\lambda$ by the elementary volume of a 
steady-state nebula is given by:
\begin{equation}
{\rm F(X^m,\lambda)} = N_{\rm e} N({\rm X^n) f(X^m,\lambda,}N_{\rm e},T_{\rm e}) \epsilon_l,
\end{equation}

where $T_{\rm e}$ is the electron temperature, $N_{\rm e}$ the electron density,  X$^n$ the ion involved in the transition, 
f(X$^m$,$\lambda$,$N_{\rm e}$,$T_{\rm e}$) the emissivity function for the line 
(X$^m$,$\lambda$) and  $\epsilon_l$ the local filling factor (fractional volume actually filled by matter with electron density $N_{\rm e}$).

The emissivity function f(X$^m$,$\lambda$,$N_{\rm e}$,$T_{\rm e}$) depends on the excitation mechanism. Recombination through the capture of an 
electron (followed by a cascade to lower  levels) and collisional excitation by an electron (followed by spontaneous radiation) are the main 
excitation processes in PNe.
$\footnote{ We overlook other possible excitation mechanisms, namely (a) Bowen resonance fluorescence (a photon emitted by one ion is absorbed
 by another and degraded in a cascade of emission lines via intermediate levels), (b) starlight and/or nebular continuum fluorescence (a strong local UV radiation 
field pumps 
an ion into an excited state), (c) dielectronic recombinations (a core electron is excited by capture of a free electron) and (d) charge-exchange reactions 
(an electron is exchanged during the 
collision of a ion with H or He, the most abundant species). Their possible role into a specific application (i. e. abundance dichotomy from optical 
recombination and collisionally excited lines) will be introduced in Sect. 5.}$

For recombination lines (n=m+1 in Eq. (1))

\begin{equation}
{\rm f(X^m,\lambda_{ij}},N_{\rm e},T_{\rm e})={\rm \frac{hc}{\lambda_{ij}}\alpha_{eff}(\lambda_{ij})},
\end{equation}
where $\alpha_{eff}(\lambda_{ij})$ is the effective line recombination coefficient (a weak function of $N_{\rm e}$) given by
\begin{equation}
{\rm \alpha_{eff}(\lambda_{ij})=B(\lambda_{ij})\alpha_{eff}(X^m_i)}
\end{equation}
and B($\lambda_{ij}$), the branching ratio, by 
\begin{equation}
{\rm B(\lambda_{ij})=\frac{A_{ij}}{\sum_{k<i} A_{ik}}},
\end{equation}
with A$_{ik}$=probabilities for spontaneous radiative decay.

The emissivity function for collisionally excited lines (low-density case; n=m in Eq. (1)) is  
\begin{equation}
{\rm f(X^m,\lambda,T_{\rm e})=\frac{hc}{\lambda} q_{coll}(\lambda)}
\end{equation}
and q$_{coll}$($\lambda$), the collisional excitation rate,  
\begin{equation}
{\rm q_{coll}(\lambda)=8.63\times 10^{-6}(\frac{\Omega}{\omega_1}) T_{\rm e}^{-0.5} e^\frac{-\Delta E}{kT_{\rm e}}},
\end{equation}
where $\Omega$=collision strength of the transition averaged over the Maxwell distribution, $\omega_1$=statistical weight of the lower level of the transition, 
$\Delta$E=excitation potential of the upper level.

In general, $\alpha_{eff}(\lambda)$ has a weak, inverse dependence on $T_{\rm e}$, whereas q$_{coll}(\lambda)$, according to Eq. (6), is a strong, direct function of 
$T_{\rm e}$.

Eq. (1) - relating the local electron density with an observable quantity (i. e. the local flux) -  represents the starting point for the practical determination 
of the radial 
matter profile in  real-PNe (after de-projection of the apparent bi-dimensional morphology, as discussed in the next section).

\section{Practice}

Let us consider an optically thick, regularly expanding nebula   
at a distance D, 
%expanding with $V_{\rm exp}$=A$\times$R'' and 
powered by a luminous post-AGB star at high temperature. 
A spectrograph long-slit centered on the exciting star intercepts the radial slice of nebula shown in the left-most panel of Fig. 1, whose high-resolution 
spectral image in 
different ions - also presented in Fig. 1 -  enhances  (a) the large stratification of the radiation and kinematics, and (b) the blurred 
appearance of H I, He I and He II recombination lines, due to a mix of thermal motions, fine-structure and expansion velocity gradient across the 
nebula (more details are in Sabbadin et al. 2005 and Sect. 4.1).\footnote {The twelve spectral images in Fig. 1 refer to NGC 6741 
at PA=15$\degr$ (ESO NTT+EMMI).} 
\begin{figure*}
   \centering \includegraphics[width=18cm]{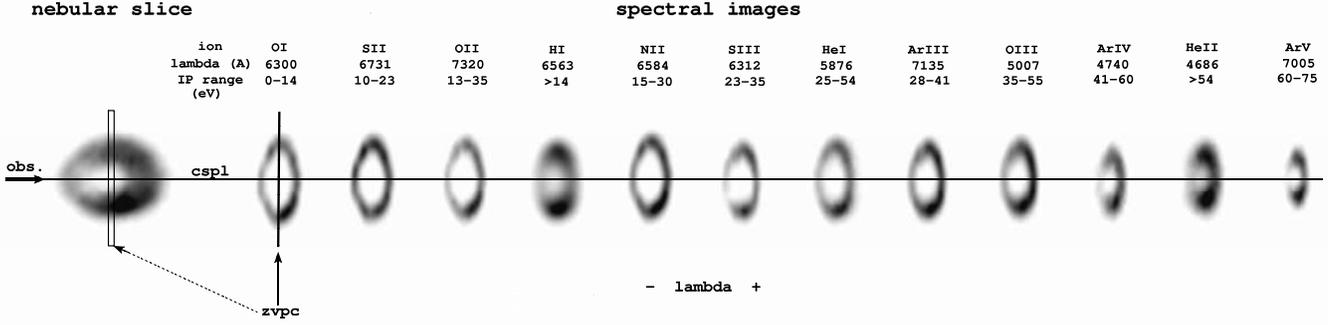}
   \caption{Nebular slice intercepted by a spectrograph slit passing through the central star (left-most panel) and the related high-resolution spectral image (flux 
in arbitrary scale) in 12 
ionic species arranged (left to right) in order of increasing ionization potential (IP; approximative range). The cspl is also indicated, as well as  
a  zvpc (superimposed to the  
[O I]  $\lambda$6300 $\rm\AA\/$ spectral image) with the corresponding part of real nebula (the rectangle centered on the nebular slice).}  
\end{figure*}

Moreover, Fig. 1 contains: 
\begin{description}
\item [-] the {\bf central-star-pixel-line} (cspl), common to all PA, representing the matter projected at the apparent position of the central 
star, whose motion is purely radial;   
\item [-] a {\bf zero-velocity-pixel-column} (zvpc, superimposed to the [O I] spectral image at $\lambda$6300 $\rm\AA\/$), giving the spatial profile 
of the tangentially 
moving gas at the systemic radial velocity.
\end{description}

Hereafter we will  focus on the zvpc: as clearly shown in the left-most panel of Fig. 1, it provides the ionic distribution within a sharp, central, 
well-defined portion of the radial slice of nebula 
selected by the spectrograph slit 
(strictly speaking, the zvpc is marginally affected by the adjacent layers of the nebular slice - and, at the same time, the zvpc marginally contributes 
to the emission of the adjacent layers - since the instrumental spectral resolution is not infinite). 

According to Eq. (1), the absolute flux emitted in the line $\lambda$ by the nebular volume sampled by a single pixel of the zvpc is given by

\begin{equation}
{\rm E(X^m,\lambda)_{zvpc} = F(X^m,\lambda)_{zvpc} \times V_{zvpc}}, 
\end{equation}
where $V_{zvpc}$ (in cm$^3$) is represented by the parallelepiped 
\begin{equation}
{\rm V_{zvpc}=s \times w \times d}
\end{equation}
with s, pixel size along the slit, given by
\begin{equation}
{\rm s(cm)= pixel\,height(arcsec)\times \frac{D(cm)}{206265}},
\end{equation}

w, pixel size perpendicular to the slit, by 
\begin{equation}
{\rm w(cm)= slit\,width(arcsec) \times \frac{D(cm)}{206265}}
\end{equation}
and d, depth of the zvpc along the radial direction, in general by 
\begin{equation}
{\rm d(cm)=\delta V(km\,s^{-1})\times\frac{R(\lambda)(arcsec)}{V_{\rm exp}(\lambda)(km\, s^{-1})}\times\frac{D(cm)}{206265}}  
\end{equation}
and, for $V_{\rm exp}$=A$\times$R'', by 
\begin{equation}
{\rm d(cm)=\frac{\delta V(km \,s^{-1})}{A(km \,s^{-1} \,arcsec^{-1})}\times \frac{D(cm)}{206265}} , 
\end{equation}
$\delta$V(km s$^{-1}$) being the pixel spectral resolution.

The corresponding flux received by an observer is 
\begin{equation}
{\rm I(X^m,\lambda)_{\rm obs}= \frac{I(X^m,\lambda)_{\rm corr}}{10^{[c(H\beta)\times f(\lambda)]}}}
\end{equation}
with 
\begin{equation}
{\rm I(X^m,\lambda)_{\rm corr}=\frac {E(X^m,\lambda)_{zvpc}}{4\pi D(cm)^2}}
\end{equation}
where c(H$\beta$) is the logarithmic extinction at H$\beta$ and f($\lambda$) the extinction coefficient given by Seaton (1979).

In the ionized nebula being $N(\rm H)\simeq$$N(\rm H^+$) and  $N_{\rm e}\simeq$1.15$N(\rm H^+$), 
Eq. (7) can be written in the form
\footnote{In this and the following equations,  the proportionality constant is [4$\pi$(206265)$^3$$\times$A(km s$^{-1}$ arcsec$^{-1}$)] / 
[1.15$\times \delta$V(km s$^{-1}$)$\times$
s(arcsec)$\times$ w(arcsec) $\times$D(cm)] 
 for $V_{\rm exp}$=A$\times$R'' and [4$\pi$(206265)$^3$$\times$$V_{\rm exp}(\lambda)$(km s$^{-1}$] / 
[1.15$\times \delta$V(km s$^{-1}$)$\times R(\lambda$)(arcsec)$\times$
s(arcsec)$\times$ w(arcsec) $\times$D(cm)] for $V_{\rm exp}\not=$ A$\times$R$\arcsec$.}
\begin{equation}
N{\rm _{\rm e}^2  \epsilon_l \propto [\frac{I(X^m,\lambda)_{\rm corr}} {f(X^m,\lambda,N{\rm _e},T_{\rm e})}]\times \frac{N(H)}{N(X)} \times \rm icf(X^n)}
\end{equation}
where  icf(X$^n$)= $\frac {N{\rm (X)}}{N{\rm (X^n)}}$ is the ionization correcting factor.

Recalling that
\begin{equation}
{\rm \sum_{i \le z} \frac{1}{\rm icf(X^i)}=1}
\end{equation}
(z=atomic number of the element X), at last we obtain the general expression
\begin{equation}
N{\rm _{\rm e}^2  \epsilon_l \propto \frac{N(H)}{N(X)} \times \sum_{i \le z} \frac{I(X^i,\lambda)_{\rm corr}}{f(X^i,\lambda,N_{\rm e},T_{\rm e})}}
\end{equation}
providing $N_{\rm e}\epsilon_l^{1/2}$ in the zvpc from observable quantities (i. e. 
the pixel-to-pixel flux in one line for each ionic species of a given element). 

In practice, for hydrogen - which has a single ionized state - Eq. (17) becomes
\begin{equation}
N{\rm _{\rm e}^2  \epsilon_l \propto [\frac{I(HI,6563)_{\rm corr}}{\alpha_{eff}(HI,6563)}\times \frac{6563}{hc}]} 
\end{equation}
or
\begin{equation}
N{\rm _{\rm e}^2  \epsilon_l\propto [\frac{I(HI,4861)_{\rm corr}}{\alpha_{eff} 
(HI,4861)}\times \frac{4861}{hc}]} 
\end{equation}
or using any other H I Balmer recombination line. 

For helium, the contribution of both ionization stages must be considered 
\begin{equation}
N{\rm _{\rm e}^2  \epsilon_l \propto [C(He^+)+C(He^{++})]\times \frac{N(H)}{N(He)}}
\end{equation} 
with 
\begin{equation}
{\rm C(He^+)=[\frac{I(HeI,5876)_{\rm corr}}{\alpha_{eff}(HeI,5876)}\times \frac{5876}{hc}]}
\end{equation}
and
\begin{equation}
{\rm C(He^{++})=[\frac{I(HeII,4686)_{\rm corr}}{\alpha_{eff} 
(HeII,4686)}\times \frac{4686}{hc}]}
\end{equation}
or using any other pair of He I and He II recombination lines.

%\begin{equation}
%N_{\rm e}^2  \epsilon_l \propto [(\frac{I(HeI,5876)_{\rm corr}}{\alpha_{eff}(HeI,5876)}\times \frac{5876}{hc}) + 
%(\frac{I(HeII,4686)_{\rm corr}}{\alpha_{eff} 
%(HeII,4686)}\times \frac{4686}{hc})]\times \frac{N(H)}{N(He)}
%\end{equation}
%(or an other pair of He I and He II nebular emissions).

For heavier elements, we are forced to select collisionally excited emissions of suitable ionic sequences, due to the weakness of recombination lines, large 
stratification of the radiation and  
incomplete ionic coverage of the echellograms. 
As tracers, we adopt O$^0$+O$^+$+O$^{++}$ for the  
external, low-to-medium ionization layers (IP range from 0 to 55 eV) and Ar$^{++}$+Ar$^{3+}$+Ar$^{4+}$ for the internal, medium-to-high ionization ones 
(IP range from 28 to 75 eV). $\footnote{Note, however, that ionic species higher than Ar$^{4+}$ are expected in the innermost layers of a PN powered by 
a central star with T$_*\ge$100\,000 K. In this case, $\lambda$7005 $\rm\AA\/$ and/or $\lambda$6435 $\rm\AA\/$ of Ar$^{4+}$ become poor density diagnostics 
for the internal, 
highest-excitation regions ($\lambda$3425  $\rm\AA\/$ of [Ne V], IP range 97 to 126 eV, should be preferable).}$ 

From oxygen
\begin{equation}
N{\rm _{\rm e}^2  \epsilon_l \propto [C(O^0)+C(O^+)+C(O^{++})]\times \frac{N(H)}{N(O)}}
\end{equation}
with
\begin{equation}
{\rm C(O^0)=[\frac{I(OI,6300)_{\rm corr}}{q_{coll}(OI,6300)}\times \frac{6300}{hc}]},
\end{equation}
\begin{equation}
{\rm C(O^+)=[\frac{I(OII,7320)_{\rm corr}}{q_{coll}(OII,7320) } 
\times \frac{7320}{hc}]}
\end{equation}
(7320 corresponding to $\lambda$7320.121 $\rm\AA\/$, the strongest component of the [O II] red quartet) 
and 
\begin{equation}
{\rm C(O^{++})=[\frac{I(OIII,5007)_{\rm corr}}{q_{coll}(OIII,5007)} 
\times \frac{5007}{hc})]}.
\end{equation}
%\begin{equation}
%N_{\rm e}^2  \epsilon_l \propto [(\frac{I(OI,6300)_{\rm corr}}{q_{coll}(OI,6300)}\times \frac{6300}{hc}) + 
%(\frac{I(OII,7320)_{\rm corr}}{q_{coll}(OII,7320)} 
%\times \frac{7320}{hc})+ 
%(\frac{I(OIII,5007)_{\rm corr}}{q_{coll}(OIII,5007)} 
%\times \frac{5007}{hc})]\times \frac{N(H)}{N(O)}.
%\end{equation}
From argon 
\begin{equation}
N{\rm _{\rm e}^2  \epsilon_l \propto [C(Ar^{++})+C(Ar^{3+})+C(Ar^{4+})]\times \frac{N(H)}{N(Ar)}}
\end{equation}
with
\begin{equation}
{\rm C(Ar^{++})=[\frac{I(ArIII,7135)_{\rm corr}}{q_{coll}(ArIII,7135)}\times \frac{7135}{hc}]},
\end{equation}
\begin{equation}
{\rm C(Ar^{3+})=[\frac{I(ArIV,4740)_{\rm corr}}{q_{coll}(ArIV,4740)} \times \frac{4740}{hc}]}
\end{equation}
and 
\begin{equation}
{\rm C(Ar^{4+})=[\frac{I(ArV,7005)_{\rm corr}}{q_{coll}(ArV,7005)} 
\times \frac{7005}{hc}]}.
\end{equation}

Besides Eqs. (17) to (30), the $N_{\rm e}$ profile in the zvpc is also given by line intensity ratios of ions in p$^3$ configuration, like 
$\lambda$6717 $\rm\AA\/$/$\lambda$6731 $\rm\AA\/$ of 
[S II] for low-ionization regions and $\lambda$4711 $\rm\AA\/$/$\lambda$4740 $\rm\AA\/$ of [Ar IV] for the high-ionization ones. However, these diagnostics 
are generally weak in PNe and $N_{\rm e}$[S II] and $N_{\rm e}$[Ar IV] can be derived only at (or close to) the corresponding intensity peak. 

Summing up: in the most favourable case (i. e. long-slit, wide spectral range, absolute flux calibrated echellograms of a PN at known distance), 
we infer both $N_{\rm e}$(zvpc) and  $\epsilon_l$(zvpc) by combining Eqs. (17) to (30) with $N_{\rm e}$[S II] and $N_{\rm e}$[Ar IV] (since 
(a) chemical abundances come 
from line fluxes integrated over the whole spatial and spectral profile, Alexander \& Balick 1997, Perinotto et al. 1998, and (b) the $T_{\rm e}$(zvpc) 
distribution comes 
from diagnostics of ions in p$^2$ and p$^4$ 
configurations, like $\lambda$6584 $\rm\AA\/$/$\lambda$5755 $\rm\AA\/$ of [N II] for low-ionization regions and   $\lambda$5007 $\rm\AA\/$/$\lambda$4363 
$\rm\AA\/$ of [O III] for the high-ionization ones, Turatto et al. 2002, Benetti et al. 2003).

On the other hand, in the most unfavourable case (i. e. flux un-calibrated spectra of a PN at un-known distance)  Eqs. (17) to (30) provide the ``relative'' 
$N_{\rm e}\epsilon_l^{1/2}$ profile in the zvpc.  \footnote {Moreover, (a) the combination of Eqs. (15) and (17) gives the 
radial ionization structure of each element (more details are in Sect. 4.1) and (b) the accurate He/H, O/H and Ar/H radial chemical profiles can be obtained 
by assembling Eqs. (18), (19), (20), (23) and (27); we will deepen this topic in a future, dedicated paper. }

In order 

- to check the general validity and applicability of our $N_{\rm e}$ reconstruction method for different physical and evolutionary conditions of the 
ionized gas and 

- to tackle the various ``practical'' problems connected with the observations of a true-PN (e. g. finite spatial and spectral resolutions and 
incomplete ionic coverage of the echellograms, thermal motions, turbulence and fine-structure of recombination lines), 

in the next section we 

a) will expose an arbitrary and imaginary model nebula to a series of UV stellar fluxes and 

b) will analyse in detail the zvpc-flux profile in a few, representative true-PNe.

\section{Theory vs. practice}

According to the most recent theoretical evolutionary models and radiation-hydrodynamics simulations, a PN is the result of the synergistic effects of 
ionization and fast wind on the gas ejected during the superwind phase of an AGB-star (Bl\"ocker \& Sch\"onberner 1990; Vassiliadis \& Wood 1994; 
Bl\"ocker 1995; Marigo et al. 2001; Perinotto et al. 2004b). 

By combining the large variation of post-AGB stellar characteristics (temperature and 
luminosity) with the gradual nebular dilution due to expansion, we can sketch the evolution of a ``normal''  PN as follows: 
in the early phases the nebula is optically thick to the UV stellar radiation, then the gas gradually becomes optically thin until the 
quick luminosity decline of the hot star at the end of the hydrogen-shell nuclear burning, leading to a recombination of the external nebular layers; later on, 
the PN enters in a final re-ionization phase, 
because of the slowing down in the stellar luminosity decline.

The recombination and re-ionization 
phases are absent in a low-mass PN ionized by a low-mass, slowly evolving central star, whereas massive PNe powered by a high-mass, fast evolving post-AGB star 
never become optically thin to the UV stellar radiation.

A further complication is represented by a possible PN ``rejuvenation'': the hydrogen-deficient post-AGB star can suffer a 
late thermal pulse during the motion towards the white-dwarf region, mixing and burning  hydrogen on a convective turn-over time scale and producing 
a ``born-again'' PN (Bl\"ocker 1995, 2001; Herwig et al. 1999).

Of course, all this represents an over-simplification, each real PN being an extremely complex and inhomogeneous structure: it can be optically thick at some radial 
directions 
(e.g. along and close to the dense equatorial regions) and thin at other directions (e. g. along the low-density polar caps). 

For illustrative  purposes, we will expose a single, ``standard'' radial matter profile (representing the gas 
distribution of an ideal, average, spherically symmetric  
nebula) to a series of UV stellar fluxes 
mimicing the evolution of a post-AGB star.  The resulting 
ionization structure of each ``model-PN'' snapshot, provided by the photo-ionization code 
CLOUDY (Ferland et al. 1998), will be analysed in detail and coupled  with the zvpc-flux profiles observed in a true-PN at a comparable evolutionary phase. 

The input parameters for the model nebula are summarized in Table 1, whereas Table 2 contains the integrated flux in the main optical emissions 
(relative to I(H$\beta$)=100), the absolute H$\beta$ flux and the resulting ionized mass at four representative evolutionary steps. 

Let us start with a ``quite young'' PN close to the thick-thin transition, ionized by a luminous post-AGB star of 0.605 
M$_\odot$ at a moderate temperature (T$_*$=50\,000 K), 
thus postponing to Sect. 4.4 the special case combining proto-PNe and low-excitation ``born again'' PNe.

\begin{table}
\caption{Input parameters for the adopted, spherically symmetric model nebula (CLOUDY)}
\begin{tabular}{ll}
\hline
\\
Gas density profile &  The same, standard and  arbitrary \\
&double-peak+halo distribution at \\
&all the four evolutionary steps\\
& (see Figs. 2, 4, 5 and 6; cfr. Sect. 4)\\
%&neutral gas = cfr. Sects. 3 and 4\\
\\
Chemical abundances   & He=-0.958, C=-3.30, N=-3.75, \\

 log[N(X)/N(H)]& O=-3.30, F=-6.522, Ne=-3.90,  \\
 & Na=-5.52, Mg=-5.30, Al=-6.568, \\
 &Si=-5.00, P=-6.70, S=-5.10,\\
 &Cl=-6.66, Ar=-5.60, K=-7.11,\\ 
 &Ca=-6.64, Fe=-6.30,\\
 & other elements= CLOUDY default\\

&\\
Dust                   & CLOUDY default\\
&\\
Local filling factor         & 1.0 \\
&\\
%&blackbody distribution\\
%Exciting star   & T$_*$= cfr. Sects 4.1 to 4.4\\
%&L$_*$ = cfr. Sects 4.1 to 4.4 \\
%&\\
%Distance& 2.0 kpc\\
%\\
\hline
\end{tabular}
\end{table}

\begin{centering}
\begin{table*}
\caption{Four-step model-PN evolution. Integrated intensity of representative nebular emissions (relative to I(H$\beta$)=100), absolute H$\beta$ flux 
and ionized mass for the adopted, spherically symmetric nebula (Table 1) powered by a hydrogen-burning post-AGB star of 0.605 
M$_\odot$.}
\begin{tabular}{cccccc}
\hline
\\
$\lambda$ ($\rm\AA\/$)&Ion &&I($\lambda$)&\\
\cline {3-6}
\\
&& {\bf Sect. 4.4}& {\bf Sect. 4.1}&{\bf Sect. 4.2} &{\bf Sect. 4.3}\\
&&{\bf proto-PN + }&{\bf early-PN}&{\bf intermediate-PN}&{\bf late-PN} \\
&&{\bf ``born-again'' low-exc.-PN}&&&\\
&&(very thick)&(nearly thin)& (thin)& (recombining) \\
&&T$_*$=25\,000 K & T$_*$=50\,000 K&T$_*$=100\,000 K     & T$_*$=150\,000 K  \\
&&L$_*$/L$_\odot$=6500& L$_*$/L$_\odot$=6500& L$_*$/L$_\odot$=6500& L$_*$/L$_\odot$=1000 \\
\\
\hline
\\
3425  &$[$Ne V$]$   & 0.0       & 0.0       & 4.1  & 14.7  \\
3726+29  &$[$O II$]$& 133.3  & 231.2       & 73.8       &   587.5   \\
3968  &$[$Ne III$]$   &  0.02      & 15.2       & 49.8  & 64.9  \\
4686  &He II   &  0.0      &  0.25      & 19.0  & 42.2  \\
4740  &$[$Ar IV$]$   &0.0        & 0.22       & 3.4  &2.9   \\
4861&H I   & 100.0       & 100.0       &100.0   & 100.0  \\
5007  &$[$O III$]$   & 5.5       &  490.0      &1510.0   &1312.2   \\
5876  &He I   &   3.11     & 15.5       & 12.6  & 10.5  \\
6300  &$[$O I$]$   & 1.3       & 2.5       &0.1   & 26.3  \\
6563  &H I   &  297.2      & 291.7       &285.7   & 288.4  \\
6584&$[$N II$]$   &265.5        &  169.0      &37.3   &434.5   \\
6717+31  &$[$S II$]$   &  43.1      & 24.4       &12.4   & 107.3  \\
7005  &$[$Ar V$]$   &  0.0      &  0.0      &0.77   & 1.05  \\
7135  &$[$Ar III$]$   & 1.6       & 17.4       & 21.0  & 30.7  \\
7320  &$[$O II$]$   & 0.8       &  2.2      &1.0   & 9.8  \\
\\
\\
log E(H$\beta$)&(erg s$^{-1}$)&34.549& 35.057&35.019&34.321 \\
\\
\\
M$_{ion}$& (M$_\odot$)&0.10&0.38&0.40&0.07$^{(*)}$\\
&&&&&(*) excluding the ionized mass \\
&&&&&of the recombining halo\\
\\
\hline
\end{tabular}
\end{table*}
\end{centering}

\subsection{Early PN evolution (Column 4 of Table 2)}
\begin{figure*}
   \centering \includegraphics[width=12cm]{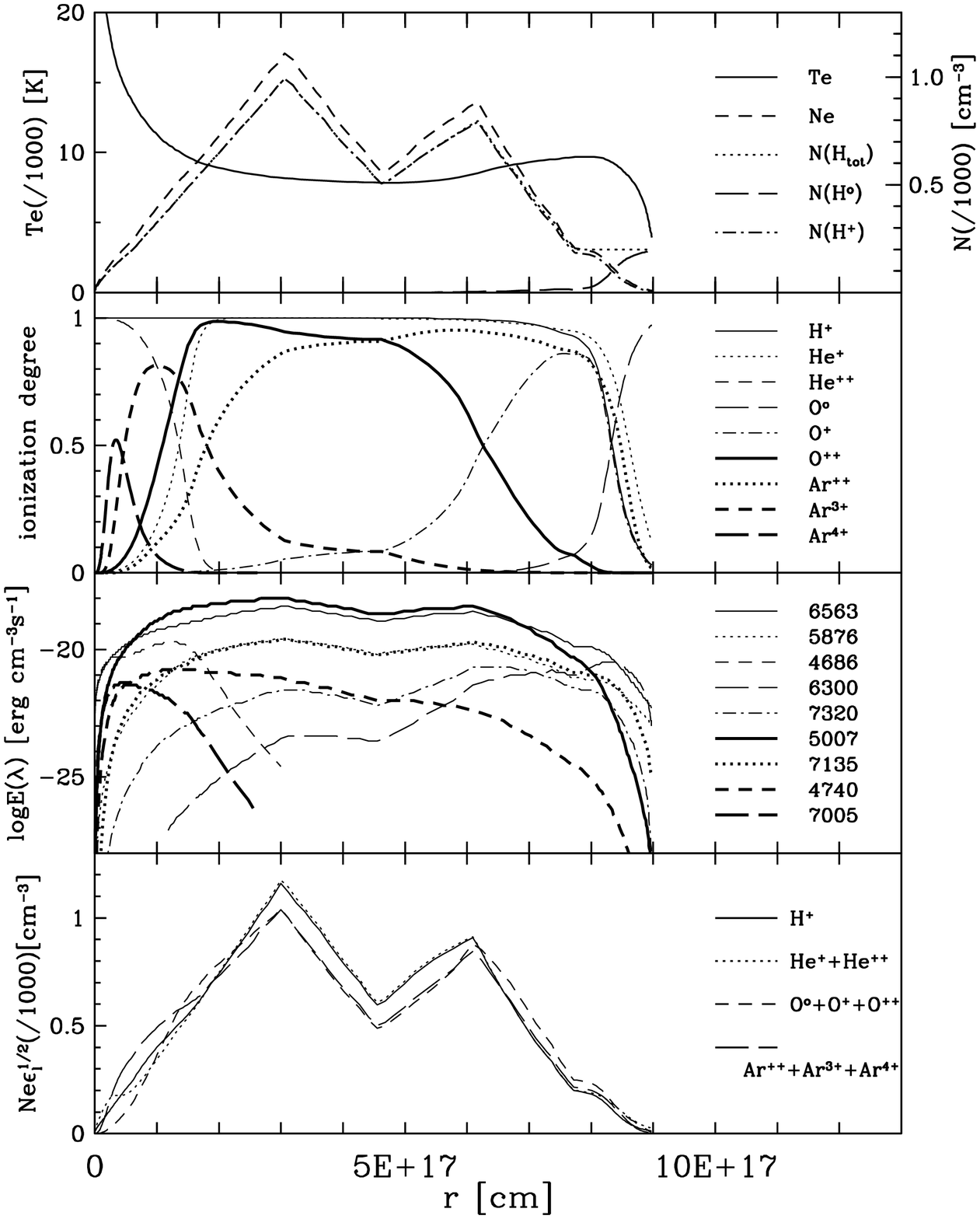}
   \caption{Early model-PN evolution (Column 4 of Table 2). General properties of the  standard, spherically symmetric model nebula (Table 1) powered by a 
central star  with  T$_*$=50\,000 K and 
L$_*$/L$_\odot$=6500. 
Top panel: adopted radial matter profile and resulting trend for the physical conditions. Next to top panel: radial ionization structure of the gas. 
Next to bottom panel: absolute flux in the main emissions. Bottom panel: $N_{\rm e}\epsilon_l^{1/2}$  distribution given by Eqs. (17) to (30), assuming 
$T_{\rm e}$=8000 K for all ions, except 
in the innermost nebular regions, where $T_{\rm e}$(He$^{++}$)=$T_{\rm e}$(Ar$^{4+}$)=13\,000 K. $N_{\rm e}\epsilon_l^{1/2}$ can be directly 
compared with 
the electron density profile shown in the top panel, since we have adopted $\epsilon_l$(model nebula)=1.}  
\end{figure*}
In this first evolutionary snapshot the ionizing engine of the adopted, spherically symmetric matter distribution (Table 1) is a hydrogen-burning post-AGB 
star of 0.605 
M$_\odot$ with  
T$_*$=50\,000 K, L$_*$/L$_\odot$=6500 and t$_{SW}$=time elapsed from the end of the superwind$\simeq$3000 yr  (Bl\"ocker 1995 and references therein). 
%\footnote{The use of a single, ``standard'' radial density profile for all snapshots can create some discrepancy between t$_{SW}$ and the true nebular age, 
%t$_{\rm PN}$. 
%Assuming  t$_{\rm PN}$$\simeq$2R$_{\rm PN}$/ [$V_{\rm exp}$(today)+$V_{\rm exp}$(AGB)],  we obtain  t$_{\rm PN}\simeq$5300 yr
%(for $V_{\rm exp}\propto$R, $V_{\rm exp}$(today)=40 km s$^{-1}$ at R$_{\rm PN}\simeq$0.15 pc and $V_{\rm exp}$(AGB)=15 km s$^{-1}$). 
%Thus, in this first evolutionary step the combination of stellar and nebular characteristics 
%corresponds to a PN powered by a hydrogen-burning post-AGB star of 0.58--0.59 M$_\odot$ (i. e. a bit lower than the average mass of PNe nuclei.) All this is 
%negligible, since it does not   
%modify at all the quantitative results obtained here and in the following sections.}

The results for the model nebula (summarized in Table 2, Column 4) are graphically shown in Fig. 2, where 
\begin{description}
\item[top panel:] standard radial matter profile and resulting trend for the physical conditions,

\item[next to top panel:] radial ionization structure of the gas,

\item[next to bottom panel:] absolute flux in the main emissions,

\item[bottom panel:] $N_{\rm e}\epsilon_l^{1/2}$  distribution given by Eqs. (17) to (30). In our case $N_{\rm e}\epsilon_l^{1/2}$=$N_{\rm e}$, since we have 
assumed $\epsilon_l$=1 
(see Table 1 ).
%(assuming $T_{\rm e}$=10\,000 K). In our case $N_{\rm e}\epsilon_l^{1/2}$=$N_{\rm e}$, since $\epsilon_l$=1 (see Table 1 ).
\end{description}
High ionization  emissions, like He II and Ar V, are very weak or absent in Fig. 2 (as expected from the  ``moderate'' star temperature), whereas the enhancement of 
[O I] and  [O II] at the extreme nebular edge  indicates that the model-PN is nearly optically thin to the UV stellar radiation. 

$T_{\rm e}$ is high ($>$10\,000 K) in the innermost, tenuous layers, rapidly decreases to $\simeq$8000 K in the densest regions, slightly 
increases further ($\simeq$9500 K) and definitively falls in the outermost, partially neutral strata. 

The bottom panel of Fig. 2 shows the $N_{\rm e}\epsilon_l^{1/2}$  profile obtained from H$^+$, He$^+$+He$^{++}$, O$^0$+O$^+$+O$^{++}$ and 
Ar$^{++}$+Ar$^{3+}$+Ar$^{4+}$ 
(through Eqs. (17) to (30) and assuming $T_{\rm e}$=8000 K for all ions, except 
in the innermost nebular regions, where $T_{\rm e}$(He$^{++}$)=$T_{\rm e}$(Ar$^{4+}$)=13\,000 K). Please note the satisfactory agreement with the input 
radial density distribution (top panel of Fig. 2),  confirming the general validity (and applicability) of the procedure developed 
in Sect. 3. \footnote{Although the adopted matter distribution of the model nebula is open to criticisms (it is arbitrary, schematic, static etc.), we underline that 
the choice of more sophisticated and/or time dependent density profiles does not modify the general results
illustrated here and in the following sub-sections.}

From theory to practice: NGC 6572 (PNG 034.6+11.8, Acker et al. 1992) is a representative example of PN in a quite early evolutionary phase. It 
consists of a 
bright (log F(H$\beta$)=-9.82 
erg cm$^{-2}$ s$^{-1}$), moderate excitation 
(excitation class 5, Hyung et al. 1994), high-density (log$N_{\rm e}$=3.7 to 4.4, Hyung et al. 1994, Liu et al. 2004) irregular disk (d$\simeq$5.0$\arcsec$) 
embedded in 
a faint envelope (10.0$\arcsec$x16.0$\arcsec$) elongated in PA$\simeq$0$\degr$.

Wilson (1950) measured $V_{\rm exp}$([O I])=16.0 km s$^{-1}$, $V_{\rm exp}$([O II])=16.85 km s$^{-1}$, $V_{\rm exp}$([S II])=15.5 km s$^{-1}$ and 
$V_{\rm exp}$([N II])=14.75 km s$^{-1}$, Weedman (1968) gave the general expansion law $V_{\rm exp}$(km s$^{-1}$)=3.4$\times$R$\arcsec$ and  Miranda et al. 
(1999, H$\alpha$+[N II] echellograms at 3 PA; $\Delta$V=8.0 km s$^{-1}$) evidenced the presence of a collimated bipolar outflow along and close to the 
apparent major axis and of an equatorial density enhancement of toroidal structure with $V_{\rm exp}$(H$\alpha$)=14.0 km s$^{-1}$ and 
$V_{\rm exp}$([N II])=18.0 km s$^{-1}$.

The exciting star of NGC 6572 has m$_B\simeq$13.0, spectral type Of/WR (Acker et al. 1992), log(T$_{\rm Z}$H I)$\simeq$log(T$_{\rm Z}$He II)$\simeq$4.83 
(Phillips 2003 and references therein) and log L$_*$/L$_\odot$=3.50($\pm0.20$) (for an assumed distance of 1200 pc, Hajian et al. 1995, Kawamura \& Masson 1996, 
Miranda et al. 1999, and c(H$\beta$)$\simeq$0.45, Hyung et al. 1994, Liu et al. 2004). It is losing mass at a large, although uncertain, rate 
(1.0$\times$10$^{-9}$ M$_\odot$ yr$^{-1}$, 6.3$\times$10$^{-9}$ M$_\odot$ yr$^{-1}$ and 3.0$\times$10$^{-8}$ M$_\odot$ yr$^{-1}$ according to Cerruti-Sola \& 
Perinotto 1985, Hutsemekers \& Surdej 1989 and Modigliani et al. 1993, respectively) and a terminal wind velocity of 1800 km s$^{-1}$.

The closeness of H I and He II Zanstra temperature of the star and the enhancement of low ionization species 
([O I], [O II] etc.) at the nebular edge suggest that NGC 6572 is still optically thick to the UV stellar radiation.

We have observed the nebula at six equally spaced PA with the high-resolution cross dispersed echelle spectrograph SARG (Gratton et al. 2001) mounted at TNG 
(Telescopio Nazionale 
Galileo), under non-photometric 
sky conditions and seeing ranging between 0.50\arcsec\,and 0.70\arcsec. The spectrograph slit (0.40\arcsec \,wide 
and 26.7\arcsec\,long) was  
centered on the exciting star. 
The echellograms (exposure time 360s) cover the spectral range  $\lambda$$\lambda$4600--8000 $\rm\AA\/$ with instrumental resolution 
$\Delta$V$\simeq$ 114\,000 (2.63 km s$^{-1}$), pixel spectral scale $\delta$V=1.20 km s$^{-1}$  and pixel spatial scale  s=0.167 arcsec. 
The spectra were reprocessed according to the straightforward procedure described by  Turatto et al. (2002), including bias, zero--order flat field and distortion 
corrections and wavelength and relative flux \footnote{Only ``relative'' 
fluxes are considered here, since the 
observations have been secured in non-photometric sky conditions. Note, however, that an absolute calibration can be obtained by means of the H$\beta$, H$\alpha$ 
and [O III] HST imagery of 
NGC 6572; in fact, the absolute flux within the 0.40$\arcsec\times$26.7$\arcsec$ rectangle of HST image intercepted by the spectrograph slit matches 
the integrated flux of the corresponding spectral emission. } calibrations. 

Let us select the spectrum at PA=102$\degr$, i. e. close to the apparent minor axis of NGC 6572, since, according to the de-projection criteria introduced 
by Sabbadin et al. (2005), R$_{\rm zvpc}$(apparent minor axis)$\simeq$ R$_{\rm cspl}$. 

The cspl vs. zvpc relation for NGC 6572 at  PA=102$\degr$ - based on ten ionic species, i.e. H$^+$, He$^+$, O$^0$, O$^+$, O$^{++}$, N$^+$, S$^+$, S$^{++}$, 
Ar$^{++}$ and Ar$^{+3}$ - gives the expansion 
law $V_{\rm exp}$(km s$^{-1}$)=8.0($\pm$1)$\times$R$\arcsec$, whose 
slope is much larger than Weedman's (1968) report. \footnote{Weedman obtained  $V_{\rm exp}$(km s$^{-1}$)=3.4$\times$R$\arcsec$ assuming 
R$_{\rm zvpc}$(apparent major axis) 
$\simeq$ R$_{\rm cspl}$.} This implies that the dynamical age of the nebula is about 750 yr and the true age 1200-1400 yr. 

Due to the nebular compactness and low expansion velocity (SS$\simeq$RR$\simeq$5, see Sect. 1), the spectral images must be carefully de-blurred. To this end we apply 
the Richardson-Lucy algorithm (Richardson 1972, Lucy 1974) 
with point-spread 
function given by a bi-dimensional Gaussian profile characterized by W$_{\rm seeing}$ and W(vel)$_{\rm tot}$, where:  
\begin{description}
\item[W$_{\rm seeing}$] - FWHM of the instrumental profile 
along  the spatial axis of the echellograms - corresponds to 0.52$\arcsec$ (3.11 pixels),  
\item[W(vel)$_{\rm tot}$] - FWHM of the broadening function 
along  the velocity axis of the echellograms - is given by  
\end{description}
\begin{equation}
{\rm W(vel)_{ tot}=[W_{\rm SARG}^2 + W_{\rm thermal}^2 + W_{\rm turb.}^2 + W_{\rm fine-s.}^2]^{1/2}}
\end{equation}
with 
\begin{description}
\item[W$_{\rm SARG}$] = instrumental resolution = $\Delta$V =2.63 km s$^{-1}$,

\item[W$_{\rm thermal}$] = thermal motions = 21.6$\times$10$^{-2}\times$$T_{\rm e}$$^{0.5}\times$ ${\rm m}$$^{-0.5}$ km s$^{-1}$ 
(m=atomic weight of the element); we assume $T_{\rm e}$=11\,000 K,

\item[W$_{\rm turb.}$] = random, small-scale motions = 5.0 km s$^{-1}$ (adopted value),

\item[W$_{\rm fine-s.}$] = fine-structure of recombination lines = 7.7 km s$^{-1}$ for H$\beta$ and H$\alpha$ and 5.0 km s$^{-1}$ for 
$\lambda$4686 $\rm\AA\/$ of He II (after suppression of the 20 km s$^{-1}$ blue-shifted tail; for details, see Sabbadin et al. 2005) 
and $\lambda$5876 $\rm\AA\/$ of He I. 
\end{description} 

W(vel)$_{\rm tot}$ amounts to $\simeq$23 km s$^{-1}$ ($\simeq$19 pixels) for H$\beta$ and H$\alpha$, $\simeq$13 km s$^{-1}$ ($\simeq$11 pixels) for 
$\lambda$5876 $\rm\AA\/$ of He I and $\lambda$4686 $\rm\AA\/$ of He II, and $\simeq$7 to 8 km s$^{-1}$ (6 to 7 pixels) for the forbidden lines of heavier 
elements (N, O, S, Ar etc.). 
\begin{figure*}
   \centering \includegraphics[width=12cm]{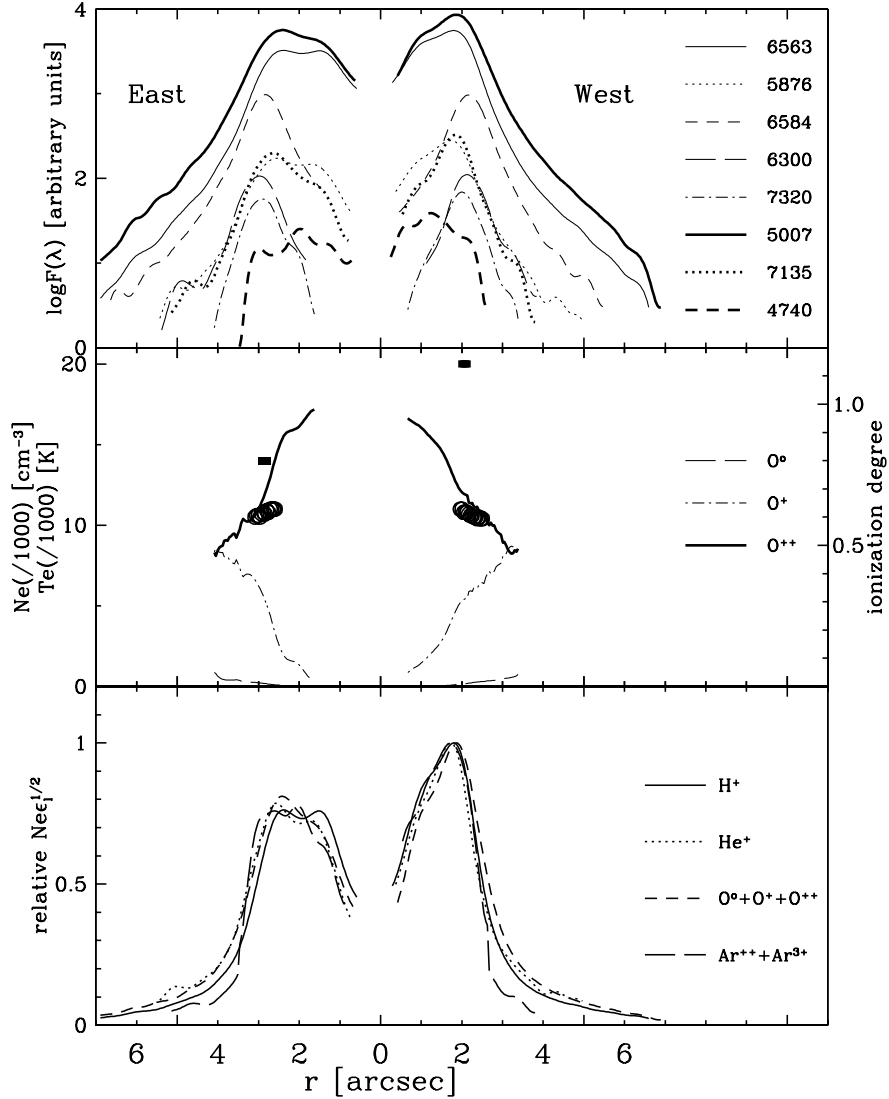}
   \caption{Early evolution of a true-PN. General properties in the zvpc of NGC 6572 at  PA=102$\degr$ (apparent minor axis). 
Top panel: flux distribution of the main nebular emissions  
across the whole nebula (in arbitrary scale and corrected for c(H$\beta$)=0.45). Middle panel: $T_{\rm e}$([N II]) 
(circles) and $N_{\rm e}$([S II]) (squares) at the corresponding line peak and ionization structure of oxygen (from Eq. (32) and 
assuming $T_{\rm e}$= 11\,000 K all across the nebula). 
Bottom panel: relative 
$N_{\rm e}\epsilon_l^{1/2}$ profile from $\lambda$6563 $\rm\AA\/$ (H$^+$), $\lambda$5876 $\rm\AA\/$ (He$^+$), $\lambda$6300 $\rm\AA\/$ (O$^0$) + $\lambda$7320 
$\rm\AA\/$ (O$^+$) + $\lambda$5007 $\rm\AA\/$ (O$^{++}$) and $\lambda$7135 $\rm\AA\/$ (Ar$^{++}$) + $\lambda$4740 $\rm\AA\/$ (Ar$^{3+}$) 
(through Eqs. (17) to (30)).  
To be noticed: the zvpc crosses the entire true-PN and provides the gas characteristics at two opposite radial directions in the plane of the sky. 
The central star position is at r=0.0 arcsec and the slit orientation is shown in the top panel.}  
\end{figure*}
The detailed results for the zvpc of NGC 6572 at PA=102$\degr$ are illustrated in Fig. 3, whose top panel contains the flux distribution 
across the whole nebula 
\footnote{ 
According to Fig. 1, the zvpc of a true-PN crosses the entire object and provides two independent density profiles at 
opposite directions (in Fig. 3 the central star position is at r=0.0 arcsec and the slit orientation is indicated in the top panel). Of course, the two zvpc 
gas distributions should coincide for a spherically symmetric nebula (as in the case of the standard model-PN adopted in this paper; that's why Fig. 2 
and the next figures concerning the model-PN show the radial density profile at a single direction).} (in arbitrary 
scale and corrected for c(H$\beta$)=0.45) of the main nebular emissions. Note the absence of the highest ionization lines: 
$\lambda$6435 $\rm\AA\/$ and $\lambda$7005 $\rm\AA\/$ of [Ar V] 
and $\lambda$6560  $\rm\AA\/$ of He II are below the detection limit, 
whereas $\lambda$4686  $\rm\AA\/$ of He II lies at the extreme blue edge of the frame and is too noisy.

The middle panel of Fig. 3 shows:

(a)  $T_{\rm e}$([N II]) (circles) and $N_{\rm e}$([S II]) (squares) at the corresponding line peak. Both diagnostics are uncertain, since the 
[N II] auroral line is quite faint and the $\lambda$6717 $\rm\AA\/$/$\lambda$6731 $\rm\AA\/$ intensity ratio 
close to the high-density limit, where collisional de-excitations dominate and the diagnostic is a weak function of $N_{\rm e}$; 

(b) the radial ionization structure of oxygen. According to Eqs. (15) and (17), it is given by
\begin{equation}
\frac{N{\rm (X^i)}}{N{\rm (X)}}= \frac{1}{{\rm icf(X^i)}}
={\rm \frac{\frac{I(X^i,\lambda)_{\rm corr}}{f(X^i,\lambda,N_{\rm e},T_{\rm e})}}{\sum_i\frac{I(X^i,\lambda)_{\rm corr}}{f(X^i,\lambda,N_{\rm e},T_{\rm e})}}}
\end{equation}
with  i=0, 1 and 2  (moreover, we take $T_{\rm e}$= 11\,000 K in the whole nebula). 
Although Eq. (32) is stricty valid in the portion of zvpc containing the emission of all the three ionic species, we can assume 
$\frac{N{\rm (O^0)}}{N{\rm (O)}}$=0 for 
I(O$^0$,6300)=0, I(O$^{+}$,7320)$>$0 and I(O$^{++}$,5007)$>$0. 
%(II) $\frac{N(Ar^{4+})}{N(Ar)}$=0 for I(Ar$^{++}$,7135)$>$0, I(Ar$^{3+}$,4740)$>$0 and I(Ar$^{4+}$,7005)=0. 
\footnote {Please note that the radial ionization 
structure given by Eq. (32) is independent on the nebular spatio-kinematics (i. e. expansion velocity field and matter distribution), since both the 
 numerator and  denumerator refer to the same pixel of the zvpc.}

Fig. 3 (bottom panel) shows the relative $N_{\rm e}\epsilon_l^{1/2}$ profile from H$^+$, He$^+$, O$^0$+O$^+$+O$^{++}$ and Ar$^{++}$+Ar$^{3+}$ 
(through Eqs. (17) to (30) and assuming $T_{\rm e}$= 11\,000 K all across the nebula). 
As expected, the ionic sequence of argon underestimates $N_{\rm e}\epsilon_l^{1/2}$ 
in the external, low ionization 
regions. Moreover,  $N_{\rm e}\epsilon_l^{1/2}$ given by hydrogen is weakened by the complex  de-blurring procedure and  
oxygen slightly underestimates $N_{\rm e}\epsilon_l^{1/2}$ in the innermost, highest ionization layers (the same is valid for helium and argon, due to the 
absence of He II and Ar V emissions in the spectrum).

Despite these minor discrepancies, the bottom panel of Fig. 3 accurately reconstructs the broad and asymmetrical  ``relative'' $N_{\rm e}\epsilon_l^{1/2}$ profile 
along the  entire true minor 
axis of  NGC 6572; when combined with $N_{\rm e}$([S II]) (middle panel), it provides $N_{\rm e}$(top)$\simeq$21\,000 cm$^{-3}$ (for $\epsilon_l$=1).

The matter distribution at the other PA confirms the inhomogeneous and elongated structure of NGC 6572, consisting of a dense equatorial torus and two 
extended and fast polar caps at lower density, embedded in a tenuous, ellipsoidal, low-excitation cocoon. The nebula is optically thick to the stellar 
radiation and the ionized mass, 
given by the observed $N_{\rm e}$ spatial distribution, the H$\beta$ flux, and 
the radio flux (Aller 1984, Pottasch 1984, Osterbrock 1989, and Turatto et al. 2002), amounts to  M$_{\rm ion}$$\simeq$0.08($\pm$0.02) 
M$_\odot$.

We end recalling that 

(a) the overall spatial structure of NGC 6572 is available at {\bf http://web.pd.astro.it/sabbadin}, showing the multi-color appearance 
and opaque 
reconstruction in [O III] and [N II] (at high, medium and low flux cuts) for a rotation around the East-West axis, 

(b) further examples of ``quite young'' PNe ionized by a luminous star at moderate temperature, 
close to the thick-thin transition are NGC 6210, NGC 6567, NGC 6803, NGC 6826 and NGC 6891.

\subsection{Intermediate PN evolution (Column 5 of Table 2)}
\begin{figure*}
   \centering \includegraphics[width=12cm]{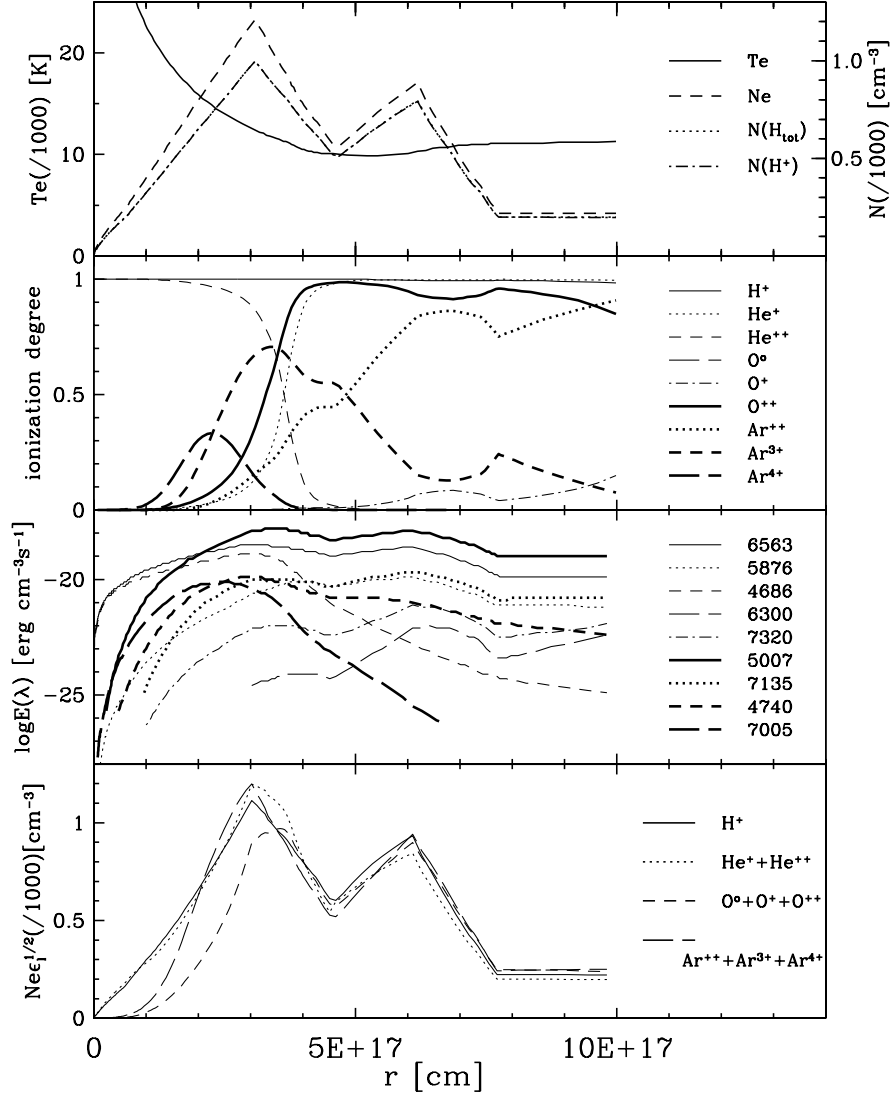}
   \caption{Intermediate model-PN evolution (Column 5 of Table 2). Radial characteristics of the usual, standard,  
spherically symmetric model nebula (Table 1) ionized by  a 0.605 
M$_\odot$ hydrogen-burning post-AGB star  with  T$_*$=100\,000 K 
and L$_*$/L$_\odot$=6500. 
Same symbology as Fig. 2.}  
\end{figure*}
In this second evolutionary snapshot  
the usual, standard, double-peak+halo  radial matter profile of the spherically symmetric model nebula (Table 1) is ionized by a 0.605 
M$_\odot$ hydrogen-burning post-AGB star  with  
T$_*$=100\,000 K, L$_*$/L$_\odot$=6500 and t$_{SW}\simeq$4900 yr  (Bl\"ocker 1995). 

Column 5 of Table 2 gives the overall properties of the model nebula, whereas the detailed radial characteristics (adopted density profile, physical conditions, 
ionization 
structure and electron density profile obtained by Eqs. (17) to (30)) are shown in Fig. 4. 
The strength of He II and [Ar V] emissions, the absence of  [O I] and weakness of [O II]  indicate that this high-excitation 
model-PN is decidedly thin to the UV stellar flux.  
$T_{\rm e}$ exceeds 20\,000 K in the innermost, low-density layers, gradually decreases to $\simeq$10\,000 K in the main regions and increases to 
$\simeq$11\,000 K further. 

Going to the bottom panel of Fig. 4, the H$\alpha$ flux, combined with the assumption $T_{\rm e}$(H$^+$)=10\,000 K all across the nebula, gives a  satisfactory 
$N_{\rm e}\epsilon_l^{1/2}$ profile; it can be further improved  adopting  distinct $T_{\rm e}$ values for the inner and the outer layers. 

The same consideration applies to 
helium lines (for $T_{\rm e}$(He$^{++}$)=15\,000 K and $T_{\rm e}$(He$^+$)=10\,000 K).

The ionic sequence of oxygen essentially reduces to O$^{++}$ (with a minor contribution of O$^+$) and closely reproduces the external, medium excitation 
regions of the model nebula, but strongly underestimates the internal, high-excitation ones (adopting $T_{\rm e}$(O$^+$)=$T_{\rm e}$(O$^{++}$)=10\,000 K).
  
Argon provides a satisfactory matter distribution across the whole nebula, except in the innermost regions, where Ar$^{5+}$ 
and even higher ionization species prevail (in Fig. 4, bottom panel, we have used  $T_{\rm e}$(Ar$^{4+}$)=15\,000 K, $T_{\rm e}$(Ar$^{3+}$)=12\,000 K and 
$T_{\rm e}$(Ar$^{++}$)=10\,000 K).

From theory to practice: NGC 7009 at PA=169$\degr$ (apparent minor axis) is the representative example of an optically thin PN powered by a luminous post-AGB 
star at high temperature, as 
shown in the extensive study by Sabbadin et al. (2004), based on the absolute flux calibrated ESO NTT+EMMI echellograms at 12 PA. We refer the reader to this 
paper for a practical application of the model-nebula discussed in this section (as well for a detailed reconstruction of the intriguing 3-D ionization 
structure of the Saturn Nebula).

More examples of optically thin, intermediate evolution PNe excited by a luminous star at high temperature are: NGC 2022, NGC 4361, NGC 6058, NGC 6804, 
NGC 7094 and NGC 7662.

We end noting that low-mass PNe ionized by a low-mass, slowly evolving central star become optically thinner and at higher excitation 
in the late evolution (thus,  the considerations developed in this section remain valid), whereas ``normal'' and/or massive PNe powered by a ``normal'' 
and/or massive star turn to a 
recombination-reionization phase (as discussed in the next section).

\subsection{Late PN evolution (Column 6 of Table 2)}
According to Bl\"ocker (1995), at time t$_{SW}\simeq$7400 yr the 0.605 M$_\odot$ hydrogen-burning post-AGB star  has T$_*$=150\,000 K and L$_*$/L$_\odot$=1000,  
i. e. it is exhausting the hydrogen-shell nuclear burning and quickly fading  along the white dwarf cooling sequence. The decreasing UV stellar flux is unable 
to fully ionize the nebular gas and the outer regions recombine, thus producing a faint halo (Tylenda 1986, Phillips 2000, Turatto et al. 2002, Sabbadin et al. 2005). 

Although the ionization and thermal structure of this model-PN are out of equilibrium, the static assumption remains - to a first approximation - valid, since 
the recombination time, 
t$_{\rm rec}$=1/($\alpha_{\rm B} \times N_{\rm e}$) ($\alpha_{\rm B}$=effective recombination coefficient),  is short: less than a century for 
hydrogen and even shorter for higher ionization species.  
\begin{figure*}
   \centering \includegraphics[width=12cm]{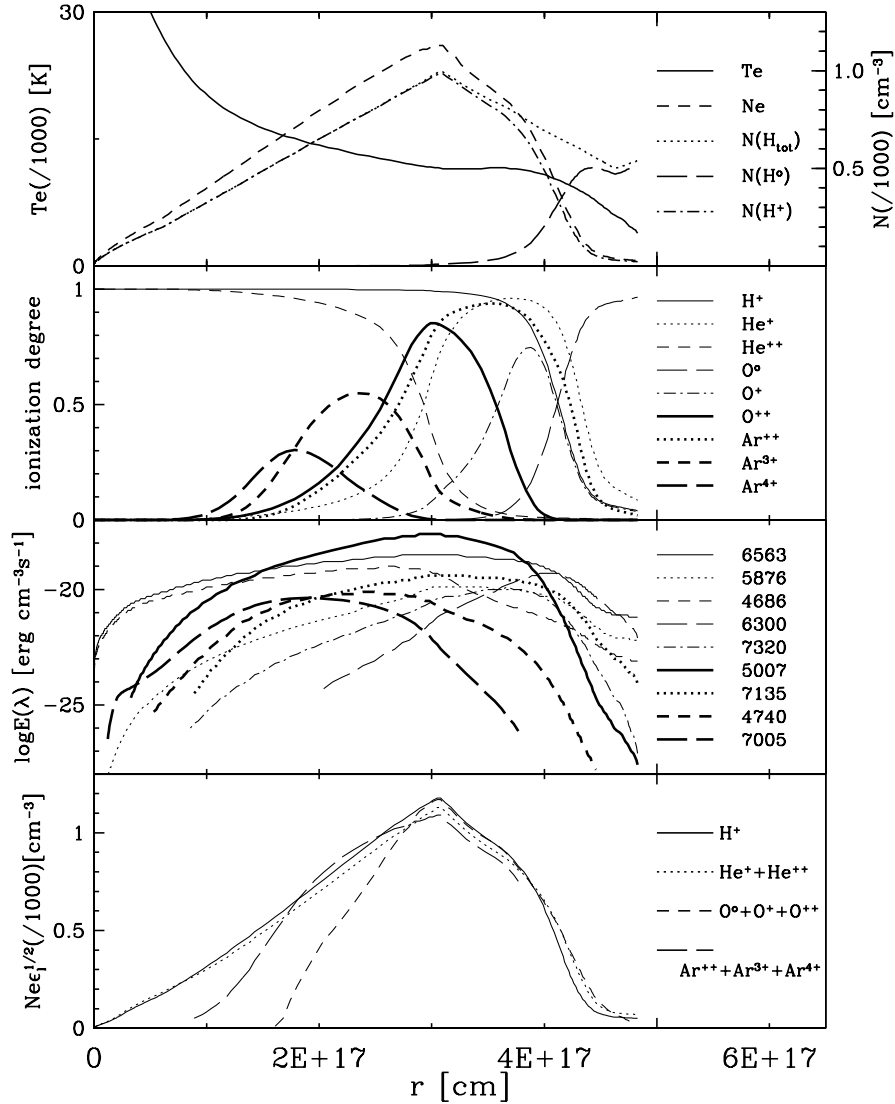}
   \caption{Late model-PN evolution (Column 6 of Table 2). General properties of the usual, standard,  
spherically symmetric model  nebula (Table 1) ionized by  a star  with  T$_*$=150\,000 K 
and L$_*$/L$_\odot$=1000. The presence of a recombining halo is ignored by the steady-state photo-ionization code CLOUDY. 
Same symbols as Figs. 2 and 4.}  
\end{figure*}
Of course, the steady-state photo-ionization code CLOUDY cannot take into account the presence of a recombining halo; this it shown in Fig. 5 - containing the radial 
characteristics of the usual, standard model nebula (Table 1) - and in Column 6 of Table 2 - giving the overall intensity in the main lines, 
the absolute H$\beta$ flux and the 
ionized mass -. 

Note (a) the very-high ionization degree of the 
innermost regions, (b) the strength of [O I] and [O II] emissions 
at the ionized edge and (c) the temperature 
radial profile: $T_{\rm e}$ exceeds 30\,000 K in the innermost, low-density layers, rapidly drops to $\simeq$11\,500 K in the densest regions 
and gradually decreases further.\footnote{Points (a) and (c) explain the central hollow normally observed in the H$\alpha$, [O III] and [N II] imagery  of 
evolved PNe, like NGC 6720 (Ring Nebula) and  NGC 7293 (Helix Nebula).}

In Fig. 5 (bottom panel) the line-fluxes of both hydrogen (for $T_{\rm e}$(H$^+$)=11\,500 K) and helium (for $T_{\rm e}$(He$^{++}$)=18\,000 K and 
$T_{\rm e}$(He$^+$)=11\,500 K) closely reproduce the $N_{\rm e}\epsilon_l^{1/2}$ model profile, whereas argon (for $T_{\rm e}$(Ar$^{4+}$)=20\,000 K, 
$T_{\rm e}$(Ar$^{3+}$)=15\,000 K and 
$T_{\rm e}$(Ar$^{++}$)=11\,500 K) underestimates the density of the innermost layers (where Ar$^{5+}$ 
and even higher ionization species dominate) and oxygen (for $T_{\rm e}$(O$^{++}$)=11\,500 K, $T_{\rm e}$(O$^+$)=9\,500 K and 
$T_{\rm e}$(O$^0$)=8\,000 K)  only agrees in the external, medium-to-low excitation 
regions of the model nebula. 

Going to practice, NGC 6565 (Turatto et al. 2002) and NGC 6741 (Sabbadin et al. 2005) are representative examples of recombining PNe; both are high-mass objects 
that never became 
optically thin to the UV radiation of the massive, fast evolving central star. The reader is referred to these papers - based on absolute flux calibrated 
ESO NTT+EMMI echellograms at 6 (NGC 6565) and 9 (NGC 6741) PA - for a practical application of the 
model-PN developed in this section. 

Following the selection criteria introduced by Sabbadin et al. (2005), further recombining PNe are: NGC 2440, NGC 2818, NGC 6302, NGC 6445, NGC 6620, NGC 6886, 
NGC 6894, NGC 7027, IC 4406 and Hu 1-2.

\subsection{A step backward for a special model killing  two birds with one stone: proto-PNe and low-excitation ``born-again'' PNe (Column 3 of Table 2)}

According to the evolutionary tracks by Bl\"ocker (1995), the combination T$_*$=25\,000 K and  L$_*$/L$_\odot$=6500 identifies a 0.605 
M$_\odot$ hydrogen-burning post-AGB star at time t$_{SW} \simeq$1600 yr, corresponding to the proto-PN phase: the newly born, dense, compact, very-low excitation 
nebula is only partially ionized by the (still modest) UV flux of the central star. Typical examples are He 2-47, He 2-131, He 2-138, Tc 1 and M 1-46.

On the other hand, the same stellar characteristics are found in very-low excitation 
objects,  like NGC 40 and BD+30$\degr$3639, powered by a low temperature and high luminosity star of late Wolf-Rayet spectral type  and widely regarded 
as ``born-again'' PNe 
(Bl\"ocker 1995, 2001; Herwig et al. 1999).

Thus, our standard model nebula (Table 1) exposed to the radiation field of a post-AGB star with T$_*$=25\,000 K and L$_*$/L$_\odot$=6500 refers to both 
the proto-PN phase 
and the low-excitation born-again PN scenario. 
\footnote{Note, however, that collisional de-excitation of forbidden lines is important for a typical, very-high density  ($N_{\rm e}\ge$ 
3$\times$10$^4$ cm$^{-3}$) proto-PN. The collisional de-excitation rate is
\begin{equation}
q_{de-ex}(\lambda)=8.63 \times 10^{-6}(\frac{\Omega}{\omega_1}) T_{\rm e}^{-0.5}, 
\end{equation}
thus, in the high-density case, Eq. (6) becomes
\begin{equation}
q_{eff}(\lambda)=8.63 \times 10^{-6}(\frac{\Omega}{\omega_1}) T_{\rm e}^{-0.5} \times (e^\frac{-\Delta E}{kT_{\rm e}}-1).
\end{equation}
}
\begin{figure*}
   \centering \includegraphics[width=12cm]{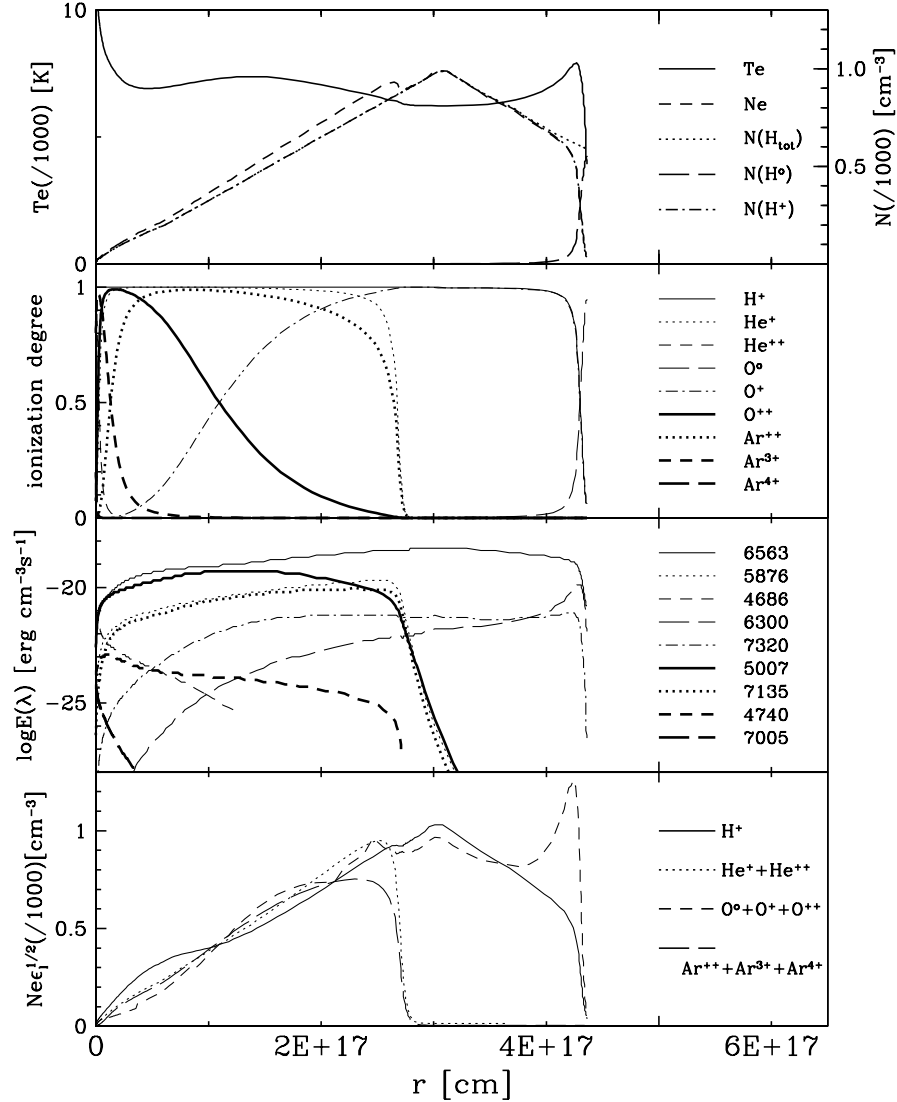}
   \caption{Model proto-PNe + low-excitation ``born-again'' PNe (Column 3 of Table 2) . Radial characteristics of the usual, standard,  
spherically 
symmetric model nebula (Table 1) for 
an ionizing   
star  with  T$_*$=25\,000 K and L$_*$/L$_\odot$=6500. 
Same symbols as Figs. 2, 4 and 5.}  
\end{figure*}

The general properties of the model nebula, summarized in Table 2 (Column 3) and graphically shown in Fig. 6, can be synthesized as follows:
\begin{description}
\item [ - ] it is optically thick (very thick!),
\item [ - ] at a very-low excitation degree (note, in particular, the weakness of $\lambda$5007 $\rm\AA\/$ ([O III]) and absence of He II, Ar IV and 
Ar V emissions),
\item [ - ] with a moderate electron temperature: only in the innermost layers $T_{\rm e}$ exceeds 8000 K, decreasing to $\simeq$6200 K in the 
densest regions (with an external bump up to 7900 K) and declining further.
\end{description}

Concerning the $N_{\rm e}\epsilon_l^{1/2}$  profile reconstructed from line fluxes through Eqs. (17) to (30) (bottom panel of Fig. 6), a satisfactory 
agreement is provided by 
H$^+$ (assuming $T_{\rm e}$(H$^+$)=6200 K), whereas  He$^+$ ($T_{\rm e}$(He$^+$)=6200 K) and Ar$^{++}$ ($T_{\rm e}$(Ar$^{++}$)=6500 K) strongly 
understimate the density of the main to external layers at low-excitation. The ionic sequence of oxygen, O$^0$+O$^+$+O$^{++}$, closely reproduces the matter 
distribution 
in the inner to main nebular regions, but presents a large discrepancy in the 
outermost parts (for ($T_{\rm e}$(O$^{++}$)=6500 K, $T_{\rm e}$(O$^+$)=6200 K and 
$T_{\rm e}$(O$^0$)=6000 K). 
This curious bump is mainly ascribable to a combination of 
\begin{description}
\item[(a)] low excitation degree of the gas 
(O$^+$ is the dominant species 
in a large fraction of the ionized nebula), 
\item[(b)] steep dependence of  $q_{coll}(OII,7320)$ on $T_{\rm e}$ (much steeper than for the other forbidden lines 
of the sample). 
\end{description}
Thus, our assumption $T_{\rm e}$(O$^+$)=6200 K, constant all across the O$^+$ zone, fails in the external regions, where $T_{\rm e}$ raises up to 
7900 K (top panel of Fig. 6): 
it overestimates N(O$^+$) (and the corresponding $N_{\rm e}$(O$^+$) contribution) by a factor of two, 
being    [$q_{coll}(7320,T_{\rm e}$=7900 K)]/ [$q_{coll}(7320,T_{\rm e}$=6200 K)]$\simeq$4.3 (whereas for the other forbidden lines of the sample 
[$q_{coll}(\lambda,T_{\rm e}$=7900 K)]/ [$q_{coll}(\lambda,T_{\rm e}$=6200 K)]=1.5 to 2.3).

Passing to practice, an illustrative example of very-low excitation PN ionized by a luminous, low temperature star  is BD+30$\degr$3639, 
covered at 
six equally spaced PA with TNG+SARG. \footnote{The spectra of BD+30$\degr$3639 and NGC 6572 have been obtained in the same (non-photometric) night and with the same 
instrumental configuration. Thus, the observative information presented in Sect. 4.1 also apply to the echellograms of BD+30$\degr$3639.} 

BD+30$\degr$3639 (Campbell's hydrogen envelope star, PNG 064.7+05.0 Acker et al. 1992) is a bright (log F(H$\beta$)=-10.03 
erg cm$^{-2}$ s$^{-1}$), high density (log$N_{\rm e}$=3.8 to 4.3, Barker 1978, Kingsburgh \& Barlow 1992, Aller \& Hyung 1995, Bernard-Salas et al. 2003),  
roundish, sharp ring (5.0$\arcsec\times$7.0$\arcsec$) surrounded by a massive envelope of neutral gas, at a distance of 1200-1500 pc (Kawamura \& Masson 
1996, Li et al. 2002). It is powered by a WC9 star with 
m$_V$=12.5,  T$_*\simeq$35\,000 K and  L$_*$/L$_\odot\simeq$4300 (Pottasch et al. 1978, K\"oppen \& Tarafdar 1978, Leuenhagen et al. 1996), 
which is losing mass at a ``moderate'' wind velocity ($\simeq$700 km s$^{-1}$) and a high - although uncertain - rate (de Freitas-Pacheco et al. 1993 
and  Leuenhagen et al. 1996 reported 6.7$\times$10$^{-6}$ M$_\odot$ yr$^{-1}$ and 1.3$\times$10$^{-5}$ M$_\odot$ yr$^{-1}$ assuming  a nebular distance of 
2.0 Kpc and 2.68 Kpc, respectively). 

Following Bryce \& Mellema (1999), BD+30$\degr$3639 presents a peculiar kinematical structure among PNe, the internal [O III] layers expanding faster than 
the external [N II] ones ($V_{\rm exp}$([O III])=35.5 km s$^{-1}$ vs. $V_{\rm exp}$([N II])=28.0 km s$^{-1}$). 
HST STIS observations in the C II] line 
at $\lambda$2327  $\rm\AA\/$ by Li et al. (2002) give $V_{\rm exp}$(C II])=36.25 km s$^{-1}$.

All this is confirmed and deepened by our TNG+SARG echellograms. In general, they indicate that:

(a) the overall excitation degree of the gas is extremely low: mean-ionization emissions (e. g. O III, He I and Ar III) are weak and high-ionization lines 
(He II, Ar IV and Ar V) totally absent,

(b) mean-ionization species move faster than the low-ionization ones,

(c) at each PA the range in R$_{\rm zvpc}$ is quite small, suggesting  a  sharp radial density profile,

(d) faint, fast-expanding knots and tails ($V_{\rm exp}$ up to 100 km s$^{-1}$) are present in the strongest emissions (as already noticed by Bryce \& Mellema 1999), 

(e) the spectral images are characterized by a series of local, small-scale irregularities and deformations. 

In detail, the cspl vs. zvpc relation at PA=10$\degr$ (apparent minor axis of BD+30$\degr$3639), based on nine ionic species,  enhances the 
presence of two distinct 
velocity fields: the external gas at low ionization (i. e. O$^0$, O$^+$, S$^+$ and N$^+$ lines) follows the  classical, Wilson's type, 
expansion law $V_{\rm exp}$(km s$^{-1}$)=15.0($\pm$2)$\times$R$\arcsec$ 
(valid for 1.875$\arcsec$$\le$R$_{zvpc}$$\le$2.10$\arcsec$), 
whereas the internal, higher ionization species (O$^{++}$, Ar$^{++}$, He$^+$ and S$^{++}$) move faster, but their velocity quickly decreases outwards 
according to the (rough) relation  
$V_{\rm exp}$(km s$^{-1}$)=-43($\pm$5)$\times$R$\arcsec$ +110($\pm$10), valid for 1.725$\arcsec$$\le$R$_{zvpc}$$\le$1.875$\arcsec$. The contact is represented 
by H$^+$, characterized by $V_{\rm exp}$(H$^+$)=28.5($\pm$1.0) km s$^{-1}$ and R$_{zvpc}$(H$^+$)=1.875($\pm$0.050) arcsec.\footnote{In other words, BD+30$\degr$3639 
is a rare case of PN exhibiting the ``U''-shaped expansion profile mentioned in Sect. 1.}

To be noticed: in the internal regions of BD+30$\degr$3639 being  $V_{\rm exp}$$\not=$A$\times$R$\arcsec$, the radial depth of the zvpc changes across the nebula 
and is given by Eq. (11), 
%\begin{equation} 
%{\rm d(\lambda) (cm)=\delta V \times \frac{R_{zvpc}(\lambda) (arcsec)}{V_{\rm exp}(\lambda) (km\, s^{-1})}\times \frac{D(cm)}{206265}}, 
%\end{equation}
where $V_{\rm exp}$($\lambda$) refers to the cspl.
\begin{figure*}
   \centering \includegraphics[width=12cm]{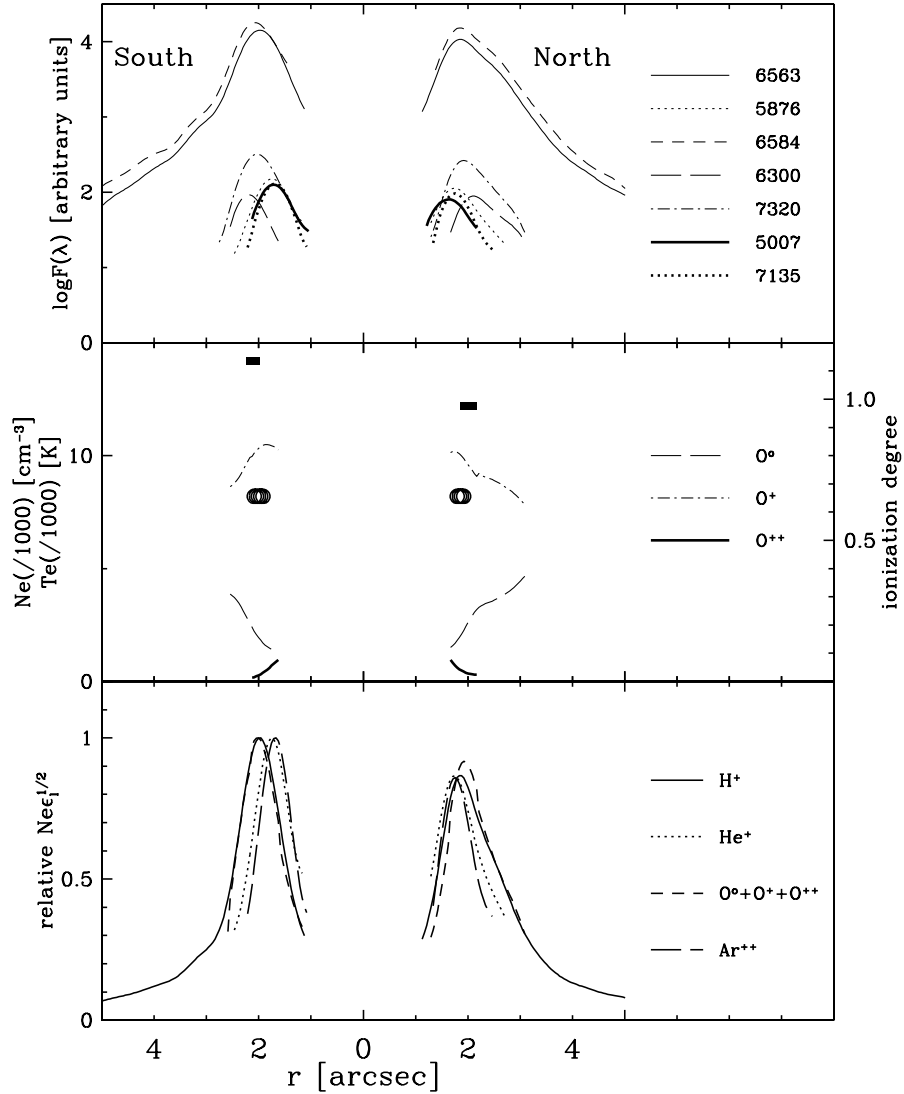}
   \caption{Real proto-PNe and low-excitation ``born-again'' PNe. General properties in the zvpc of  BD+30$\degr$3639 
at PA=10$\degr$ (apparent minor axis). Top panel: flux distribution of the main nebular emissions  
across the whole nebula (in arbitrary scale and corrected for c(H$\beta$)=0.45). 
Middle panel: $T_{\rm e}$([N II]) (circles) 
and $N_{\rm e}$([S II]) (squares) at the corresponding line peak and ionization structure of oxygen (from Eq. (32)). Bottom panel: relative 
$N_{\rm e}\epsilon_l^{1/2}$ profile from H$^+$, He$^+$, O$^0$+O$^+$+O$^{++}$ and Ar$^{++}$ 
(through Eqs. (17) to (30)). 
As in the case of NGC 6572 (Fig. 3), the zvpc provides two independent radial profiles of BD+30$\degr$3639 at opposite directions. The central star position 
is ar r=0.0 arcsec and the slit orientation 
is shown in the top panel.
Same symbols as Fig. 3.}  
\end{figure*}

The general results for the zvpc of BD+30$\degr$3639 
at PA=10$\degr$ are shown in Fig. 7,  containing the line fluxes 
across the whole nebula (in arbitrary scale and corrected for c(H$\beta$)=0.45; top panel), the physical conditions ($T_{\rm e}$([N II]) and $N_{\rm e}$([S II]) 
at the corresponding 
flux peak) and the ionization structure of oxygen (through Eq. (32) and assuming  $\frac{N(O^{++})}{N(O)}$=0 for 
I(O$^0$,6300)$>$0, I(O$^+$,7320)$>$0 and I(O$^{++}$,5007)=0; middle panel) and the relative $N_{\rm e}\epsilon_l^{1/2}$  profile (
through Eqs. (17) to (30); bottom panel). As expected, 
H$\alpha$ (for $T_{\rm e}$(H$^+$)=8200 K) and 
O$^0$+O$^+$+O$^{++}$ (for $T_{\rm e}$(O$^0$)=8000 K, $T_{\rm e}$(O$^+$)=8200 K and $T_{\rm e}$(O$^{++}$)=8500 K) provide the density distribution across the whole 
ionized nebula, whereas He$^+$ (for $T_{\rm e}$(He$^+$)=8500 K) and Ar$^{++}$ (for $T_{\rm e}$(Ar$^{++}$)=8500 K) peak in the internal regions at  
higher ionization.  

When combined with $N_{\rm e}$([S II]), the sharp matter profile in Fig. 7, bottom panel, has $N_{\rm e}$(top)$\simeq$ 15\,000 cm$^{-3}$ (for $\epsilon_l$=1) and the 
ionized mass of the nebula results to be M$_{\rm ion}$$\simeq$0.05($\pm$0.01) 
M$_\odot$. 

Although a deep analysis of BD+30$\degr$3639 (whose multi-color and opaque movies are shown in the WEB page {\bf http://web.pd.astro.it/sabbadin}) is beyond the 
aims of the paper, we 
wish to mention that:
\begin{description}
\item[ - ] the faint halo embedding the bright nebula (see the bottom panel of Fig. 7)  
could be a ``recombining'' halo, since it essentially retains the kinematical properties observed in the external layers of the 
main nebula (vice versa, the kinematics of an AGB-halo is independent on the main nebula, simply reflecting the ejection law in the pre-superwind phase; for 
details, see Sabbadin et al. 2005). If this is the case, 
the past UV flux of the exciting star was much higher than the present value; less than 200 years ago, the star abruptly cooled and/or dimmed and 
 recombination became the  
dominant process in the outermost nebular regions, no longer reached by stellar radiation;  

\item[ - ] the combination of peculiar expansion velocity field + sharpness of the gas radial profile joins 
BD+30$\degr$3639 with NGC 40, a second very-low 
excitation PN powered by a low temperature and high luminosity star of late Wolf-Rayet spectral type (Sabbadin et al. 2000), 
and identifies a small sub-class of PNe whose internal layers are likely shaped by wind interaction.  
\end{description}
\section{Discussion}
Given a regularly expanding mass of ionized gas, the zero-velocity-pixel-column (zvpc) of high-resolution spectral images represents the tangentially moving matter 
at the systemic velocity and  
identifies a sharp radial portion of nebula in the plane of the sky, whose characteristics (density, temperature and ionic  profiles) are  obtainable from 
line fluxes. \footnote {Please, note the affinity between the zvpc given by high-resolution slit spectroscopy and the rest frame of 
imaging Fabry-Perot interferometry.}

The one-dimensional distribution provided by the zvpc at each PA - combined with the kinematical properties given by the central-star-pixel-line (cspl) - 
represents the starting 
point for 
bi-dimensional tomography, i. e. recovery of the ionized gas structure 
within the entire slice of nebula intercepted by the spectrograph slit.

In his turn, bi-dimensional tomography is the basis for the spatial reconstruction of the whole nebula, through a 3-D 
rendering program assembling the tomographic maps at different PA. 

This is, in nuce, the complete logical path of our original methodology - valid for all types of expanding nebulae -  overcoming any misleading camouflage due to 
projection and providing 
the true distribution of the kinematics, physical conditions and ionic and chemical abundances at unprecedented accuracy. 

In this paper we have extracted the precious information stored in the zvpc of high-resolution spectral images of PNe, focusing on the gas density profile given by 
different atomic species. The results for  hydrogen, helium and heavier elements can be synthesized as follows:
\begin{description}
\item[ (I) ] hydrogen (the most abundant element) presents a mono-electron atomic structure with a low ionization potential (IP=12.598 eV); 
at first glance, the recombination 
lines of H I (whose emissivity is a weak function of the electron temperature) are excellent diagnostics of the matter distribution, since 
they are emitted across the whole ionized nebula. However, their effectiveness and reliability are reduced by three severe blurring sources -  
thermal motions, fine-structure and expansion velocity gradient within the PN - implying the accurate spectral image-reconstruction illustrated in Sect. 4.1;

\item[ (II) ] the same general considerations apply to the recombination lines of  both ionization stages of helium;

\item[ (III) ] for heavier elements, we are forced to adopt collisionally excited emissions of suitable ionic sequences, because of the 
faintness of recombination lines, 
large stratification of the radiation across the nebula and   incomplete ionic coverage of the echellograms.  We select O$^0$+O$^+$+O$^{++}$ for  
low-to-medium ionization layers (IP range from 0 to 55 eV) 
and Ar$^{++}$+Ar$^{3+}$+Ar$^{4+}$ for the medium-to-high ionization ones 
(IP range from 28 to 75 eV). The main weaknesses of this choice (and the corresponding care) are: (1) collisional excitation rates of  forbidden lines present 
a strong 
dependence on the electron temperature (in many cases, even a rough, ad hoc $T_{\rm e}$ profile is preferable to the assumption $T_{\rm e}$=constant across 
the whole nebula) and 
(2) ionization species higher than Ar$^{4+}$ dominate in the internal regions of high-excitation PNe 
powered by a hot, T$_*\ge$100\,000 K, central star ($\lambda$3425  $\rm\AA\/$ of [Ne V] , IP range 97.1 to 126.2 eV, could be a better diagnostic for the innermost 
layers). 
\end{description}
The illustrative examples presented in the previous sections show that the ``best observational'' density profile is always a 
compromise among the 
results given by different elements, taking into 
account the general characteristics of each target (evolutionary phase, optical thickness - or thinness - and excitation degree of the nebula and 
temperature and luminosity of the central star). 

All this, combined with a high-quality (i. e. RR$\ge$5 and SS$\ge$5) spectroscopic material and a careful reduction procedure - and thanks to the 
moderate dependence of $N_{\rm e}$ on the observational parameters (see Eqs. (17) to (30)) -, for the first time provides 
the accurate recovery (to within 10 to 20\%) of the radial matter distribution into an expanding, ionized mass of gas.

Our zvpc analysis 
\begin{description}
\item[(A)] is based on the pixel-to-pixel measurement of  velocity and flux in all nebular emissions at high, medium and low excitation present in 
long-slit echellograms covering a wide  spectral range. This strongly differs from (i) the usual observational setup for extended objects, 
inserting an interference filter to isolate a single echelle order and (ii)  the normally adopted reduction procedure, aiming either at the kinematics or at average 
line fluxes; 
\item[(B)] is limited to the optical region (i.e. the  ionized gas); for a complete understanding of the 
PN-phenomenology, it should be extended to other spectral domains, like  near-IR (for the outermost envelope, rich in molecular and photo-dissociation emissions) 
and UV 
(for very-high ionization species expected in the innermost layers, where the star-nebula interaction occurs);
\item[(C)] applies to other classes of expanding nebulae (e. g. Nova and Supernova Remnants, shells around Population I Wolf-Rayet stars, 
nebulae ejected by Symbiotic Stars, bubbles surrounding early spectral-type Main Sequence stars etc.) covered at adequate ``relative''  spatial (SS) and 
spectral (RR) resolutions (SS=R/$\Delta$r, R=apparent radius, 
$\Delta$r=seeing+guiding; RR=$V_{\rm exp}$/$\Delta$V, $\Delta$V=instrumental spectral resolution).
\end{description} 
The zvpc methodology developed in this paper - added to tomography and 3-D recovery (see Ragazzoni et al. 2001, Turatto et al. 2002, Benetti et al. 2003, 
Sabbadin et al. 2004, 2005) - overcomes the long-standing de-projection hurdle and opens new unusual ways to deepen many facets of the nebular research, in particular:
\begin{description}
\item[ (I) ] to solve the topical problem of PNe -  i.e. formation, shape and shaping - by comparing the observational results for a 
representative sample of targets at different evolutionary phases with the expectations coming from current theoretical evolutionary models (Sch\"onberner et al. 
1997, Steffen et al. 1998, Marigo et al. 2001), 
detailed hydro-dynamical simulations (Icke et al. 1992, Frank 1994, Mellema 1997, Perinotto et al. 2004b) and updated photo-ionization codes 
(Ferland et al. 1998, Ercolano et al. 2003).  
At present, the most striking discrepancies between model-PNe (Perinotto et al. 2004b) and true-PNe (Sabbadin et al. 2005 and references 
therein) concern the radial distribution of

(a) kinematics: Wilson's law is reversed in model-PNe (i. e. the internal layers of the bright main shell expand faster than the medium-to-external ones),

(b) matter: in model-PNe the sharp density peak of the main shell abruptly falls inwards, whereas in true-PNe the whole density profile and the inwards decline 
are smooth. 

We aim at overcoming these discrepancies (both qualitatively ascribable to an overestimation of the shaping effects due to wind interaction) by adjusting the 
theoretical evolutive parameters until model-PNe agree with true-PNe;

\item[ (II) ] to mine the nature and phenomenology of ``exotic'' structures at large and small scales, 
like haloes, caps, ansae, jets, blobs etc.. An emblematic case is represented by the FLIERs (fast, 
low-ionization 
emitting regions) of PNe; the acronym, introduced by Balick (1978), identifies a series of external, N overabundant  knots at low excitation and peculiar 
kinematics. Although the origin and evolution of FLIERs have been interpreted in many fanciful ways (instability zones at the ionization front, 
condensations accelerated by 
photo-evaporation or wind-interaction, high-speed stellar or nebular bullets, magnetically focused knots etc.; Perinotto et al. 2004a and references therein), 
our high-resolution spectral survey  
shows that most FLIERs are normal condensations within the outer envelope (same kinematics and chemical abundances), whose spectral characteristics are fully 
explainable in terms of mere photo-ionization by the central star;$\footnote {For a deep analysis of the FLIERs, ansae and caps of NGC 7009, see 
Sabbadin et al. (2004).}$
  
\item[ (III) ] to disentangle the source (or sources) of general, still unsolved problems involving emission nebulae, like the heavy-element abundance and 
electron temperature discrepancies 
obtained from optical recombination and collisionally excited lines. The proposed solutions (Peimbert et al. 2004, 
Liu et al. 2004, Ercolano et al. 2004, Sharpee et al. 2004 and references therein) include  (a) strong temperature and density 
fluctuations within a chemically homogeneous nebula, (b) high-density, low-temperature knots 
overabundant in heavy elements embedded in a hydrogen-rich medium at lower-density and higher-temperature and (c) other physical processes (they are listed 
in footnote 1) competing with single 
electron capture (i. e. recombination) in exciting high-level lines. 

All these (and other) hypotheses can be deeply tested by 
 comparing  the detailed spatio-kinematics of recombination and forbidden lines in different ions provided by the 
zvpc analysis, tomography and 3-D recovery 
applied to very-deep echellograms. 
\end{description}

\section{Conclusions} 
The presence of an expanding nebular gas is the typical signature of the instability phases in 
stellar evolution, characterized by a large, prolungated  mass-loss rate (PNe, Symbiotic Star nebulae, shells around Wolf-Rayet and 
LBV (luminous blue variable) stars), or even an explosive event (Nova and Supernova remnants). 

Mass-loss of evolved stars occupies a strategic 
ground between stellar and interstellar physics: it raises fundamental astrophysical problems (e. g. origin and structure of winds, 
formation and evolution of dust, synthesis of complex molecules), plays a decisive role in the final stages of stellar evolution 
and is crucial for galactic enrichment in light and heavy elements.

So far, image de-projection represented an unsourmontable obstacle for the accurate recovery of the spatial structure in real nebulae, 
thus impeding any detailed comparative analysis with theoretical models. Such a frustrating situation is now overcome by 
tomography and 3-D recovery developed at the Astronomical Observatory of Padua (Italy). 

The radial density reconstruction  illustrated in this paper represents the umpteenth proof that  
properly secured, reduced and analysed spectra at high-resolution constitute a peerless tool for deeping the spatio-kinematics, 
physical conditions, ionic structure  and evolution of all classes of expanding nebulae, thus 
bridging the present, wide  and pernicious gap between  model- and true-nebulae.

\begin{acknowledgements} We wish to thank the support staff of NTT and TNG 
(in particular, Gloria Andreuzzi, Olivier Hainaut, Antonio Magazz\'{u}, Luca Di Fabrizio e 
Walter Boschin) for the excellent assistance during 
the observations. F. S. greatly  enjoyed general discussions with Detlef Sch\"onberner.
%We would like to thank Denise Gon\c calves (the referee), Vincent Icke, Antonio Mampaso, Detlef Sch\"onberner and Noam Soker for their 
%suggestions, criticisms and encouragements, and 
%the support staff at the NTT (in particular, Olivier Hainaut) for assisting with the observations. 
%This paper has been partially financied by the grant 
%Cofin MM02905817 of the Italian Ministry of Education (MIUR) and partially 
%supported by the grant ASI (Agenzia Spaziale Italiana) I/R/70/00.
\end{acknowledgements}

%\Online{prova}

\end{document}